\documentclass[10pt,journal]{IEEEtran}
\usepackage{cite}
\usepackage{graphicx}
\usepackage{psfrag}

\usepackage[tbtags]{amsmath} 
\usepackage{amssymb}  
\usepackage{verbatim} 

\newcounter{actr}
{\begin{list}{(\alph{actr})}{\usecounter{actr}}}{\end{list}}

\newcounter{ictr}
{\begin{list}{(\roman{ictr})}{\usecounter{ictr}}}{\end{list}}

\newtheorem{thm}{Theorem}
\newtheorem{lemma}{Lemma}
\newtheorem{claim}{Claim}
\newtheorem{corol}{Corollary}

\newtheorem{defn}{Definition}

\newcommand{\defeq}{\stackrel{\Delta}{=}}

\newcommand{\mrm}{\mathrm}
\newcommand{\ip}[2]{\left\langle{#1},{#2}\right\rangle}

\newcommand{\cA}{{\mathcal{A}}}

\newcommand{\cB}{{\mathcal{B}}}

\newcommand{\bC}{{\mathbf{C}}}
\newcommand{\cC}{{\mathcal{C}}}

\newcommand{\bI}{{\mathbf{I}}}

\newcommand{\cN}{{\mathcal{N}}}

\newcommand{\cS}{{\mathcal{S}}}

\newcommand{\cU}{{\mathcal{U}}}

\newcommand{\bv}{{\mathbf{v}}}

\newcommand{\cW}{{\mathcal{W}}}

\newcommand{\cX}{{\mathcal{X}}}

\newcommand{\cY}{{\mathcal{Y}}}

\newcommand{\bY}{{\mathbf{Y}}}

\newcommand{\eps}{\varepsilon}

\usepackage[mathcal]{euscript}  

\title{Carbon Copying Onto Dirty Paper}
\author{Ashish~Khisti,~\IEEEmembership{Student~Member,~IEEE,}
        Uri~Erez,~\IEEEmembership{Member,~IEEE,}\\
        Amos~Lapidoth,~\IEEEmembership{Fellow,~IEEE,}
        Gregory~W.~Wornell~\IEEEmembership{Fellow,~IEEE}\thanks{This
        work has been supported in part by the National Science
        Foundation under Grant No.~CCF-0515109, and by Hewlett-Packard
        through the MIT/HP Alliance.  This work was presented in part
        at the International Symposium on Information Theory, Chicago,
        IL, June 2004 and the International Zurich Seminar, February 2006.}
        \thanks{A.\ Khisti and G.\ W.\ Wornell are
        with the Dept. Electrical Engineering and Computer Science,
        Massachusetts Institute of Technology, Cambridge, MA, 02139,
        USA (E-mail: \{khisti,gww\}@mit.edu).  U.\ Erez is with the
        Department of Electrical Engineering-Systems, Tel Aviv
        University, Ramat Aviv, 69978, Israel (E-mail:
        uri@eng.tau.ac.il). A.\ Lapidoth is with the Institute for
        Information and Signal Processing, Swiss Federal Institute of
        Technology (ETH) -- Zurich, CH-8092, Switzerland (E-mail:
        lapidoth@isi.ee.ethz.ch).}}

\begin{document}

\maketitle

\begin{abstract}
A generalization of the problem of writing on dirty paper is
considered in which one transmitter sends a common message to
multiple receivers.  Each receiver experiences on its link an
additive interference (in addition to the additive noise), which
is known noncausally to the transmitter but not to any of the
receivers. Applications range from wireless multi-antenna
multicasting to robust dirty paper coding.

We develop results for memoryless channels in Gaussian and binary
special cases.  In most cases, we observe that the availability of
side information at the transmitter increases capacity relative to
systems without such side information, and that the lack of side
information at the receivers decreases capacity relative to
systems with such side information.

For the noiseless binary case, we establish the capacity when
there are two receivers.  When there are many receivers, we show
that the transmitter side information provides a vanishingly small
benefit. When the interference is large and independent across the
users, we show that time sharing is optimal.

For the Gaussian case we present a coding scheme and establish its
optimality in the high signal-to-interference-plus-noise limit
when there are two receivers. When the interference is large and
independent across users we show that time-sharing is again
optimal. Connections to the problem of robust dirty paper coding
are also discussed.

\end{abstract}



%


\section{Introduction}
\label{sec:intro}

The study of communication over channels controlled by a random state
parameter known only to the transmitter was initiated by Shannon
\cite{Shannon58}.  Shannon considered the case where the state
sequence is known causally at the encoder.  Subsequently, Gel'fand and
Pinsker \cite{GelfandPinsker80} analyzed the case where the state
sequence is available noncausally.  The noncausal model has found
application in diverse areas, ranging from coding for memory with
defects \cite{KuznetsovTsybakov74,HeegardElGamal83}, to digital
watermarking \cite{chenWornell01,Cohen02, Moulin99}, and to coding for
the multiple-input/multiple-output (MIMO) broadcast channel
\cite{CaireShamai00,WSS04}.

Costa \cite{costa1983} considered a version of the Gel'fand-Pinsker
model in which there is an additive white Gaussian interference
(``dirt''), which constitutes the state, in addition to independent
additive white Gaussian noise.  The key result in this ``dirty paper
coding'' scenario is that there is no loss in capacity if the
interference is known only to the transmitter.


By contrast, there has been very limited work to date on
\emph{multiuser} channels with state parameters known to the
transmitter(s).  In an early work in this area, Gel'fand and
Pinsker \cite{GelfandPinsker84} show that the Gaussian broadcast
channel with \emph{independent messages} incurs no loss in
capacity if the interference sequences are known noncausally to
the transmitter. Some other multiuser settings are also discussed.
The degraded broadcast channel with independent messages and state
sequence known to the transmitter either causally or non-causally
is examined in \cite{Steinberg03}. Other works on multiuser
channels with state parameters include~\cite{kotagiri04},
\cite{cemalSteinberg05},\cite{YoungHanArakStyrmir04},\cite{Jafar05}
and \cite{SigurjonssonKim05}.

This paper examines the \emph{common-message} broadcast channel,
which we refer to as the \emph{multicast} channel.  Specifically,
we consider a scenario in which one transmitter broadcasts a
common message to multiple receivers.  In addition to additive
noise, associated with the link to each receiver is a
corresponding additive interference. The collection of such
interferences is thus the (random) state of the multiuser channel.
In our model, the transmitter has perfect noncausal knowledge of
all these interference sequences, but none of the receivers have
knowledge of any of them. This model and its generalizations arise
in a variety of multi-antenna wireless multicasting problems as
well as in applications of robust dirty paper coding where only
imperfect knowledge of the state is available to the transmitter.

The capacity of some binary versions of such multicast channels is
reported in \cite{khisti_thesis},\cite{Ashishisit04}.  For more
general channels, \cite{SteinbergShamai05} reports achievable
rates for broadcasting common and independent messages over a
discrete memoryless channel with noncausal state knowledge at the
transmitter. The case of two-user Gaussian channels with jointly
and individually independent identically distributed (i.i.d.)
Gaussian interferences on each link is also considered in
\cite{SteinbergShamai05}, for which it is conjectured that in the
limit of large interference, time-sharing between the two
receivers is optimum even when both are only interested in a
common message.  Among other results, in this paper we establish
that this conjecture is true. We upper bound the capacity of the
Gaussian channel and show that it approaches the time-sharing rate
in this limit. In addition, we also present a coding scheme that
is asymptotically optimal in the limit of high
signal-to-interference-plus-noise (SINR) ratio
\footnote{Throughout this work, symbol refers to a \emph{real}
symbol.}.
 %

An outline of the paper is as follows.  Section~\ref{sec:Ch-Mod}
presents the general multicast channel model of interest.  The
binary special cases of interest are analyzed in
Section~\ref{sec-Binary}, and the Gaussian special cases of
interest are analyzed in Section~\ref{sec-Gauss}.  Finally,
Section~\ref{sec-Conclusion} contains some conclusions and
directions for future work. The proofs of the converses are
deferred to the Appendices.

\section{Multicast Channel Model}
\label{sec:Ch-Mod}

%

The $K$-user multicast channel of interest is defined as follows.
\begin{defn}
\label{def:channel} A $K$-user discrete memoryless multicast
channel with random parameters consists of an input alphabet
$\cX$, output alphabets $\cY_1,\cY_2,\dots,\cY_K$ for receivers
$1,2,\dots,K$, respectively, and a state alphabet $\cS$.  For a
given state sequence
$s^n=(s_1,s_2,\dots,s_n)$ such that $s_i \in \cS$ and input
$x^n=(x_1,x_2,\dots,x_n)$ such that $x_i\in\cX$, the channel
outputs are distributed according to
\begin{equation}
p(y_1^n,y_2^n,\dots,y_K^n|x^n,s^n) = \prod_{i=1}^n p(y_{1i},y_{2i},\dots
y_{Ki}|x_i,s_i)
\end{equation}
where $y_k^n=(y_{k1},y_{k2},\dots,y_{kn})$, for all $y_{ki} \in
\cY_k$, $k=1,2,\dots,K$.  Moreover, $p(s^n) = \prod_i p(s_i)$. The
particular realization $s^n$ is known noncausally to the transmitter
before using the channel, but not to any of the $K$ receivers.
\end{defn}

It is worth emphasizing that the above definition includes the
case where the channel of User $k$ is controlled by its own state
$s_k^n$. In such cases, the joint state is, with slight abuse of
notation, $s^n=(s_1^n,s_2^n,\dots,s_K^n)$, so that
$p(s_i)=p(s_{1i},s_{2i},\dots,s_{Ki})$.

The capacity of the channel of Definition~\ref{def:channel} is defined
as follows.
\begin{defn}
\label{def:cap} A $(2^{nR},n)$ code consists of a message set
$\cW_n = \{1,2,\dots 2^{nR}\}$, an encoder $f_n:\cW_n \times \cS^n
\rightarrow \cX^n$, and $K$ decoders $g_{k,n}: \cY_k^n \rightarrow
\cW_n$ for $k=1,\ldots,K$.  The rate $R$ is \emph{achievable} if
there exists a sequence of codes such that for $W$ uniformly
distributed over $\cW_n$ we have
\begin{equation}
\lim_{n\rightarrow\infty} P_e^{n} =
\lim_{n\rightarrow\infty}\Pr\left\{\bigcup_{k=1}^K \{
g_{k,n}(Y_k^n)\neq W \}\right\} = 0. \label{eq:errprob}
\end{equation}
Note that the error probability in \eqref{eq:errprob} is averaged over
all state sequences and messages.  The capacity $C$ is the supremum of
achievable rates.
\end{defn}

In the remainder of the paper, we focus on special cases of the
memoryless channel in Definition~\ref{def:channel}.  In particular, we
focus on binary and Gaussian cases in which the state is an additive
interference; for results on the memory with defects multicast
channel, see, e.g., \cite{khisti_thesis}.

\section{Noiseless Binary Case}
\label{sec-Binary}

We first consider the noiseless binary special case of
Definition~\ref{def:channel}.  Specifically, the channel outputs
$Y_1^n,Y_2^n,\dots,Y_K^n$ depend on the input $X^n$ and the states
$S_1^n,S_2^n,\dots,S_K^n$ according to
\begin{equation}
Y_k^n = X^n \oplus S_k^n
\label{eq:binary_channel_model}
\end{equation}
where $X_i,S_{ki}\in\{0,1\}$, and $\oplus$ denotes
symbol-by-symbol modulo-two addition (i.e., exclusive-or).  In
\eqref{eq:binary_channel_model}, the memoryless case of interest
corresponds to the requirement that the
$(S_{1i},S_{2i},\dots,S_{Ki})$ for $i=1,2,\dots,n$ form an i.i.d.\
sequence of $K$-tuples.  In particular, for each $i$ the variables
$\{S_{1i}, S_{2i}, \dots, S_{Ki}\}$ may in general be
statistically dependent, and do not need to be identically
distributed.  As a result, we express our results in terms of the
properties of a generic $K$-tuple in this sequence, which we
denote by $(S_1,S_2,\dots,S_K)$.

Note that with only a single receiver ($K=1$), the capacity is
trivially 1 [\mbox{bit per channel use}],\footnote{From now on,
except in the case of ambiguity, the units of ``bits per channel
use" will be omitted.} which is achieved by interference
precancellation, i.e., by choosing $X^n=S^n\oplus B^n$, so that
$Y^n=B^n$, where $B^n$ is the bit representation for the message
$W$.  As we will now develop, when there are multiple receivers,
capacity is generally less than this ideal single-user rate.

\subsection{The Case of $K=2$ Receivers}

\begin{figure}[tbp]
\begin{center}
\psfrag{&0}{$W$}
\psfrag{&1}{Encoder}
\psfrag{&3}{$X^n$}
\psfrag{&4a}{Decoder 1}
\psfrag{&4b}{Decoder 2}
\psfrag{&5a}{$Y_1^n$}
\psfrag{&5b}{$Y_2^n$}
\psfrag{&6a}{$\hat{W}_1$}
\psfrag{&6b}{$\hat{W}_2$}
\psfrag{&7a}{$S_1^n$}
\psfrag{&7b}{$S_2^n$}
\includegraphics[width=3.25in]{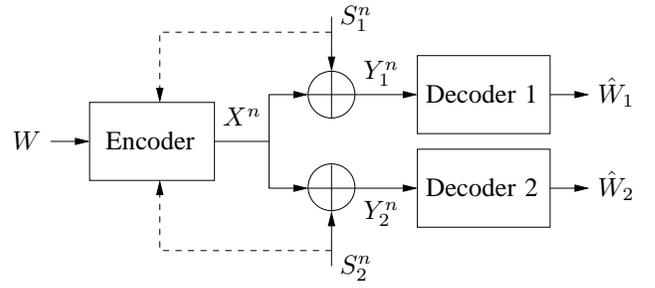}
\caption{Two-user memoryless, noiseless binary multicast channel
with additive interference.  The encoder maps message $W$ into
codeword $X^n$.  The state takes the form of interference
sequences $S_1^n$ and $S_2^n$.   Each channel output
$Y_k^n=X^n\oplus S_k^n$, where $\oplus$ denotes symbol-by-symbol
modulo-two addition, is decoded to produce message estimate
$\hat{W}_k$.} \label{fig:noiseless}
\end{center}
\end{figure}

The case of two receivers, which is depicted in
Fig.~\ref{fig:noiseless}, is the simplest nontrivial scenario since
perfect interference precancellation is not possible simultaneously
for both users.

One lower bound on the two-user capacity corresponds to a
time-sharing approach that precancels the interference of one of
the receivers at a time, yielding a rate of $R_{\mathrm{TS}}=1/2$.
Another lower bound corresponds to ignoring the interference at
the transmitter, i.e., treating each of the channels as a binary
symmetric channel.  This strategy yields a rate of
$R_{\mathrm{IS}} = 1 - \max\{H(S_1),H(S_2)\}$. It turns out that
the former bound is only tight when $S_1$ and $S_2$ are
independent and $\cB(1/2)$, and the latter bound is only tight
when both $S_1$ and $S_2$ are $\cB(0)$\footnote{We use $\cB(q)$ to
denote a Bernoulli random variable with parameter $q$ i.e.
$\Pr(S=1)=q, \Pr(S=0)=1-q$.}.

A coding theorem for the channel is as follows.
\begin{thm}
The capacity of two-user noiseless, memoryless binary channel with
additive interference is given by
\begin{equation}
C = 1 - \frac{1}{2} H(S_1 \oplus S_2).
\label{eq:2usrCapacity}
\end{equation}
\label{thm:2usrbinary}
\end{thm}

\begin{proof}
A converse is provided in Appendix~\ref{app:a}.  The achievability
argument is detailed below:

\begin{enumerate}

\item Select $2^{nR}$ codewords randomly according to an i.i.d.\
$\cB(1/2)$ distribution in a codebook $\cC$ of rate $R$ strictly
less than the capacity \eqref{eq:2usrCapacity}. Denote these
codewords as $B^n(1),B^n(2),\ldots, B^n(2^{nR})$, so a message $w$
is represented by codeword $B^n(w)$.


\item Select a sequence $A^n$ by flipping a fair coin for each
symbol index (the realization of which is also known at the
decoders \cite{WolfowitzBook}).  Select the set $\cA_1$ of symbol
indices where $A_i = 1$, and precancel the interference at those
indices for user~1, and precancel the interference at the
remaining indices $\cA_2$ (with $A_i=0$) for user~2. Specifically,
the transmitted sequence is of the form
\begin{equation}
\label{eq:binary_scheme} X_i(w) = \begin{cases}
           B_i(w)\oplus S_{1i} & i\in\cA_1 \\
       B_i(w)\oplus S_{2i} & i\in\cA_2.
      \end{cases}
\end{equation}

\end{enumerate}

With this encoding, receiver~1 then observes a version of $B^n(w)$
where $|\cA_1|$ symbols are correct, and the remaining $|\cA_2|$
symbols are corrupted by interference $S_{1i}\oplus S_{2i}$,
$i\in\cA_2$, corresponding to a binary symmetric channel with
crossover probability $q'=\Pr\{S_1\oplus S_2=1\}$.  Receiver~2
experiences the opposite effect.  Thus for large $n$ we have,
since $|\cA_1|/n\rightarrow1/2$,
\begin{equation}
\frac{1}{n} I(B^n; Y_k^n | A^n) \rightarrow \frac{1}{2} +
\frac{1}{2}(1-H(S_1 \oplus S_2)),\quad k=1,2,
\label{eq:bin_decoding_MI}
\end{equation}
which is $C$ in \eqref{eq:2usrCapacity}. As the mutual information
expression in~\eqref{eq:bin_decoding_MI} indicates, the decoding
of $Y_k^n$ to the message $\hat{W}_k$ is done by using the
knowledge of $\cA_1$ and $\cA_2$ (i.e., $A^n$) at the decoders. In
particular, receiver 1 selects a codeword which agrees with the
received symbols in the set $\cA_1$ and which is typical with
noise $S_1 \oplus S_2$ with the symbols in the set $\cA_2$. For
decoder 2, the order of the sets is reversed. As long as $R \leq
C$, $\hat{W}_k$ equals $W$ with high probability.
\end{proof}

\begin{figure}[tbp]
\begin{center}
\includegraphics[width=3.25in]{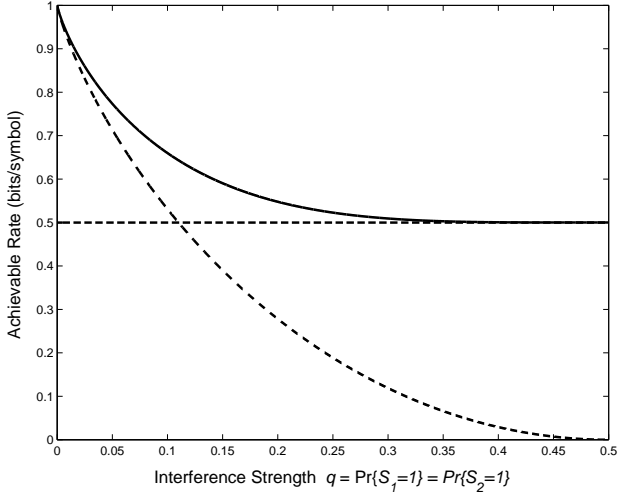}
\caption{Achievable rates for the two-user noiseless binary
multicast channel with independent and identically distributed
interferences, as a function of the strength of the interference.
Capacity is indicated by the solid curve, time-sharing performance
is indicated by the horizontal dashed line, and the performance of
a system that ignores the side information is indicated by the
downward sloping dashed curve.\label{fig:bincap}}
\end{center}
\end{figure}

Fig.~\ref{fig:bincap} shows the performance gains of optimal
coding relative to time-sharing and disregarding the side-information.
In particular, the achievable rate in the case of independent
interferences is plotted as a function of the strength of the
interference as measured by $q=\Pr\{S_1=1\}=\Pr\{S_2=1\}$.

Three immediate conclusions can be drawn from
Theorem~\ref{thm:2usrbinary}. First, transmitter-only side
information incurs a penalty relative to system-wide side
information unless $S_1$ and $S_2$ are completely dependent random
variables, i.e., unless $S_2 = S_1$ or $S_2 = \overline{S_1}$.
Second, time-sharing is strictly sub-optimal except when $S_1$ and
$S_2$ are independent $\cB(1/2)$ random variables. We emphasize
that, by contrast, when there are \emph{independent} messages for
each of the receivers in Fig.~\ref{fig:noiseless}, time-sharing
between the receivers is optimal and there is no loss in the
capacity region with side information only at the transmitter.
Finally ignoring the side information at the transmitter is
strictly suboptimal except when $H(S_1) = H(S_2) = 0$.

We make a few additional observations.

\subsubsection*{Some Further Remarks}

\begin{enumerate}

\item The achievability argument can also be obtained via a
different, more direct, but perhaps less intuitive route as
follows. First note that a straightforward extension of the random
binning argument for the single user case \cite{GelfandPinsker80}
shows that the following rate is achievable for the $K$-user
multicast channel with random parameters.
\begin{equation}
\label{eq:commonGP} R_K =  \max_{p(U|S),p(X|U,S)}\{\min_k
I(U;Y_k)-I(U;S)\},
\end{equation}
Here $U$ is an auxiliary random variable (over some alphabet
$\cU$) that satisfies the Markov constraint $U \leftrightarrow
(X,S) \leftrightarrow Y_k$ for $k=1,2,\dots,K$.

For the two-user binary channel, the following choice of $U$
yields the achievability of~\eqref{eq:2usrCapacity}. Let the
alphabet of  $U$ be $\cU =\{\Psi_1,\Psi_2,\Psi_3,\Psi_4\}$.
\begin{equation}
\label{eq:Uchoice}
\begin{aligned}
U &= A\, \{ \Psi_1\,(X\oplus S_1) + \Psi_2\,(\overline{X\oplus S_1}) \} \\
&\quad + \bar{A}\, \{ \Psi_3\, (X\oplus S_2) + \Psi_4\,
\overline{(X \oplus
  S_2)} \},
\end{aligned}
\end{equation}
where, $X$ is $\cB(1/2)$ random variable, independent of $S_1$ and
$S_2$, and $A$ is also $\cB(1/2)$ that is independent of $X$,
$S_1$ and $S_2$, and where $\bar{\cdot}$ denotes the complement of
a (binary-valued) variable.

\item For the code construction outlined above suggests the
transmitter does not require noncausal knowledge of the
interference. We emphasize, however, this result is specific to
the noiseless binary channel model.

\item It is straightforward to verify that random linear codes are
sufficient to achieve the capacity of Theorem~\ref{thm:2usrbinary}.
It suffices to use an argument analogous to that used by Gallager
for the binary symmetric channel \cite[Sec.~6.2]{Gallager68}.

\item Theorem~\ref{thm:2usrbinary} can be readily generalized to the
case of state sequences that are not in general i.i.d.  In this case
the term $H(S_1\oplus S_2)$ in \eqref{eq:2usrCapacity} is simply
replaced with the entropy \emph{rate} of $S_1^n\oplus S_2^n$.

\item Our achievability scheme also applies in the presence of
noise. For the channel model
\begin{equation*}
\begin{aligned}
Y_1 &= X \oplus S_1  \oplus Z_1\\
Y_2 &= X \oplus S_2 \oplus Z_2,
\end{aligned}
\end{equation*}
where $Z_1$ and $Z_2$ are mutually independent and identically
distributed Bernoulli random variables and independent of all
other variables, we can show that a rate
\begin{equation*}
R = 1-\frac{1}{2}H(S_1 \oplus S_2 \oplus Z_1) - \frac{1}{2}H(Z_1)
\end{equation*}
is achievable and an upper bound is given by
\begin{equation*}
R^+ = 1-\frac{1}{2}H(S_1 \oplus S_2) - \frac{1}{2}H(Z_1)
\end{equation*}
Note that time-sharing is optimal in the special case when $S_1$
and $S_2$ are independent $\cB(1/2)$ random variables.

\end{enumerate}

\subsection{The Case of $K>2$ Receivers}

\label{sec:GeneralBinary}

When there are more than two receivers  further losses in capacity
ensue, as we now develop.  Specifically, we have the following
bounds on capacity.
\begin{thm}
The capacity of the $K$-user noiseless binary channel in which the
generic $S_1, S_2, \dots, S_K$ are mutually independent and
identically distributed\footnote{Our results actually hold more
generally provided the distribution across the interference sequences
is symmetric, i.e., if for all $m$, $p(s_{k_1}, s_{k_2}, \dots,
s_{k_m})$ is independent of the specific choice of $k_1, k_2, \dots,
k_m \in \{1,2,\dots, K\}$.} is bounded according to:
\begin{subequations}
\begin{equation}
R_- \le C \le R_+,
\end{equation}
where
\begin{align}
R_+ &= 1 - \frac{1}{K}H(S_1\oplus S_2, S_1 \oplus S_3, \ldots,
S_1\oplus S_K), \label{eq:bub} \\
R_- &= \max\left\{1-H(S_1), 1 - \left(1-\frac{1}{K}\right) H(S_1 \oplus S_2)\right\}. \label{eq:blb}
\end{align}%
\label{eq:binbounds}%
\end{subequations}%
\label{thm:kuserBinary}
\end{thm}

\begin{proof}
The upper bound \eqref{eq:bub} is established in Appendix~\ref{app:b}.
The lower bound \eqref{eq:blb} is obtained via a direct generalization
of the code construction \eqref{eq:binary_scheme} in the case of two
users. Specifically, it suffices to consider a code construction that
divides each codeword into $K$ equally sized blocks and precancels
the interference for a different user in each of the blocks.  Each
user then experiences one clean block and $K-1$ noisy blocks governed
by a binary symmetric channel with crossover probability $q' =
\Pr\{S_1 \oplus S_2=1\}$ as before.
\end{proof}

\begin{figure}[tbp]
\begin{center}
\includegraphics[width=3.25in]{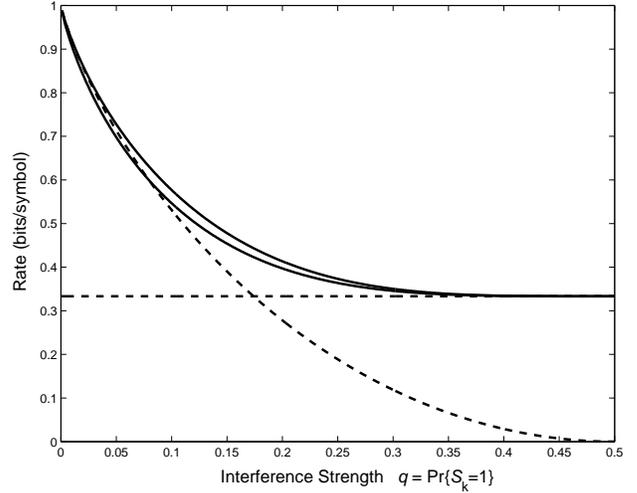}
\caption{Upper bound and lower bounds on the capacity of the
three-user noiseless binary multicast channel, as a function of the
strength of the interference.  The solid curves depict the two bounds
of \eqref{eq:binbounds}. The horizontal dashed line indicates the
performance of time-sharing, while the other dashed curve indicates
the performance of a strategy in which the side information is ignored
by the transmitter.}
\label{fig:binbounds}
\end{center}
\end{figure}

In general, the lower and upper bounds in \eqref{eq:binbounds} do not
coincide.\footnote{A slightly improved lower bound appears in
\cite{khisti_thesis}, but it, too, does not match the upper
bound.}  However, the associated rate gap decreases monotonically
with the number of receivers $K$.  Moreover, even for $K=3$, it is
small, as Fig.~\ref{fig:binbounds} illustrates.

The rate gap also decays to zero in the limit of large $K$, which
follows readily from Theorem~\ref{thm:kuserBinary}.  In particular, $C
\rightarrow 1-H(S)$ as $K\rightarrow \infty$, where $S$ denotes a
generic random variable with the distribution of the $S_k$.  To see
this, it suffices to recognize that when $S_1, S_2, \dots, S_K$ are
i.i.d.,
\begin{equation}
\begin{aligned}
\left(1-\frac{1}{K}\right)H(S) &\leq \frac{1}{K}H(S_1 \oplus
S_2, S_1 \oplus S_3, \dots, S_1 \oplus S_K)\\
 &\leq H(S).
\end{aligned}
\label{eq:rategap}
\end{equation}
As $K \rightarrow \infty$, the lower and upper bounds in
\eqref{eq:rategap} converge, so that the upper bound on capacity
\eqref{eq:bub} converges to $R_+=1-H(S)$.  However, this rate is
achievable by simply treating the interference as noise at the
receivers, so it is the limiting capacity.  It should be
emphasized that this implies that when the number of receivers is
large, the side-information available to the transmitter is
essentially useless.

We can also use \eqref{eq:rategap} to bound the rate penalty
associated with ignoring side information as a function of the
number of receivers $K$. In particular, the gap is at most
$H(S)/K$.

Finally, we can use Theorem~\ref{thm:kuserBinary} to establish
that in the limit of large interference, time-sharing is optimal
for every $K$.  Specifically, when $S_k\sim\cB(1/2)$, the capacity
is $C=1/K$ and is achieved through time-sharing.  To see this, it
suffices to specialize the upper bound in \eqref{eq:bub}.
Specifically, $S_1\oplus S_k$ for $k=2,3,\dots,K$ are independent
$\cB(1/2)$ random variables, so the joint entropy is $K-1$.

\section{Gaussian Case}
\label{sec-Gauss}

\begin{figure}[tbp]
\begin{center}
\psfrag{&0}{$W$}
\psfrag{&1}{Encoder}
\psfrag{&3}{$X^n$}
\psfrag{&4a}{Decoder 1}
\psfrag{&4b}{Decoder 2}
\psfrag{&5a}{$Y_1^n$}
\psfrag{&5b}{$Y_2^n$}
\psfrag{&6a}{$\hat{W}_1$}
\psfrag{&6b}{$\hat{W}_2$}
\psfrag{&7a}{$S_1^n$}
\psfrag{&7b}{$S_2^n$}
\psfrag{&8a}{$Z_1^n$}
\psfrag{&8b}{$Z_2^n$}
\includegraphics[width=3.25in]{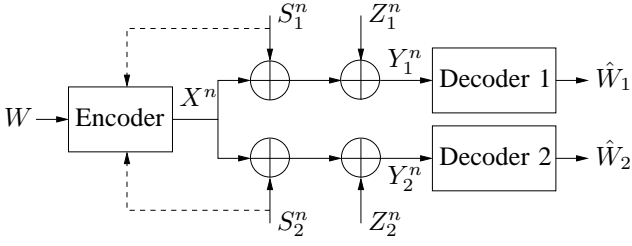}
\caption{Two-user Gaussian multicast channel model with additive
interference.  The encoder maps message $W$ into codeword $X^n$.
The state takes the form of interference sequences $S_1^n$ and
$S_2^n$. Each channel output $Y_k^n=X^n+S_k^n+Z_k^n$ is decoded to
produce message estimate $\hat{W}_k$. \label{fig:gaussian}. The
interference and noise sequences are i.i.d.\ and mutually
independent. Furthermore, $S_1,S_2\sim\cN(0,Q)$ and
$Z_1,Z_2\sim\cN(0,1)$.}
\end{center}
\end{figure}

In this section we consider a memoryless Gaussian extension of
Definition~\ref{def:channel} and incorporate an average power
constraint on the input.  Unless otherwise stated, we restrict to
the two-user ($K=2$) case.  In the scenario of interest, depicted
in Fig.~\ref{fig:gaussian}, the state is additive, and the
associated interferences $S_k^n$ are zero-mean white Gaussian
sequences of power $Q$.  We first focus on the case of independent
interferences and consider the case of correlated interferences in
section~\ref{sec:rdpc}. In addition, each receiver's link also has
a zero-mean additive white Gaussian noise $Z_k^n$ of power $N$.
Thus, the observation at receiver $k$ takes the form
\begin{equation}
Y_k^n = X^n + S_k^n + Z_k^n,\qquad k=1,2.
\label{eq:gauss-channel}
\end{equation}
Our power constraint takes the form
\begin{equation}
\label{eq:powerConstraint} \frac{1}{n} E\left[\sum_{i=1}^n
X_i^2(W,S_1^n,S_2^n)\right] \le P,
\end{equation}
where the expectation is taken over the ensemble of messages and
interference sequences. Finally, note that without loss of
generality, we may set $N=1$, and interpret $P$ as the
signal-to-noise ratio (SNR), and $Q$ as the interference-to-noise
ratio (INR).

For this channel, we present the following bounds on the capacity.
\begin{thm}
\label{thm:gaussbounds} An upper bound on the Gaussian multicast
channel capacity is :
\begin{equation}
C \le \min\{R_+^\mrm{I}, R_+^\mrm{II}\}, \label{eq:gaussbounds}
\end{equation}
where\footnote{All logarithms are to the base 2 in this work. Also
the notation $[f]^+$ refers to $\max(f,0)$ in~\eqref{eq:gub_d} and
throughout the paper.} $^,$\footnote{The trivial upper bound of
$\frac{1}{2}\log(1+P)$ is sometimes tighter than these two bounds,
particular in the limit of very small $P$. }
\begin{equation}
R_+^\mrm{I} =
\begin{cases}
\frac{1}{4}\log(1+P) + \frac{1}{4}\log\left(\frac{P+Q+1+
2\sqrt{PQ}}{Q}\right) & Q \geq 4\\
\frac{1}{4}\log\left(\frac{1+P}{Q/4+1}\right)+
\frac{1}{4}\log\left(\frac{P+Q+1+ 2\sqrt{PQ}}{Q/4+1}\right) &
 Q < 4
\end{cases}
\label{eq:gub_s}
\end{equation}
\begin{equation}
R_+^\mrm{II} =
\begin{cases}
\frac{1}{2}\log\left(\frac{1+P + Q + 2\sqrt{PQ}}{1+Q/2} \right)
& Q \leq 2\\
\frac{1}{2}\log\left(\frac{1+P+Q + 2\sqrt{PQ}}{\sqrt{2Q}}\right)-
\left[\frac{1}{4}\log\left(\frac{Q}{2P+2}\right)\right]^+ &
 Q >2
\end{cases}
\label{eq:gub_d}
\end{equation}

\end{thm}

We have presented two different upper bounds denoted by
$R_+^\mrm{II}$ and $R_+^\mrm{I}$ since neither bound dominates the
other, over all values of $(P,Q)$. The two bounds have been
derived by slightly different methods. The bound $R_+^\mrm{I}$ is
obtained by observing that the channel is non-trivial even if we
set one of the interferences (say $S_1$) to 0. Furthermore, it is
possible to show that an upper bound on this modified channel is
also an upper bound on the Gaussian multicast channel of interest.
A complete derivation of this upper bound is presented in
Appendix~\ref{app:d}. The expression for $R_+^\mrm{II}$ is
obtained by directly applying a chain of inequalities on the
Gaussian multicast channel and its derivation is presented in
Appendix~\ref{app:c}.

We remark here that the upper bounds are explicit expressions of
the following maximization:
\begin{equation}
\begin{aligned}
R_+^\mrm{I} &= \min_{\rho \in
[-1,1]}\frac{1}{4}\log\left(\frac{1+P}{1+\rho}\right)  +
\frac{1}{4}\log\left(\frac{P+Q+1+
2\sqrt{PQ}}{Q/2+1-\rho}\right)\\
\end{aligned}
\label{eq:gub_s_opt}
\end{equation}

\begin{equation}
\begin{aligned}
R_+^\mrm{II} &= \min_{\rho \in [-1,1]}
\frac{1}{2}\log\left(\frac{P+Q+2\sqrt{PQ}+1}{\sqrt{(1+\rho)(Q+1-\rho)}}\right)\\
&\quad-\left[\frac{1}{4}\log\left(\frac{Q}{2P+(1+\rho)}\right)\right]^+
\end{aligned}
\label{eq:gub_d_opt}
\end{equation}

\begin{thm}
A lower bound on the Gaussian multicast channel capacity is :
\label{thm:gauss_lower_bounds}
\begin{equation}
\label{eq:glb-exp} R_- =
\begin{cases}
\frac{1}{2}\log\left(1+\frac{P}{Q/2+1}\right) & Q/2 < 1\\
\frac{1}{2}\log\left(\frac{P+Q/2+1}{Q}\right) +
\frac{1}{4}\log\left(\frac{Q}{2}\right) & 1 \leq Q/2 < P + 1
\\
\frac{1}{4}\log(1+P) & Q/2 \geq P+1.
\end{cases}
\end{equation}

\end{thm}

\begin{proof}

The lower bound\footnote{Our lower bound for $Q/2 < 1$ was also
independently reported by Costa~\cite{CostaPrivate}.}
\eqref{eq:glb-exp} is an explicit expression of the following
maximization:

\begin{subequations}
\begin{equation}
R_- = \max_{\{(P_A,P_D): P_A\ge 0, P_D\ge 0, P_A+P_D\le P\}}
R(P_A,P_D) \label{eq:glb}
\end{equation}
with
\begin{equation}
 R(P_A,P_D) \defeq \frac{1}{2}\log\left(1 +
\frac{P_A}{P_D+Q/2+1}\right) + \frac{1}{4}\log\left(1 +
P_D\right). \label{eq:RSV}
\end{equation}
\end{subequations}

Accordingly, we show the achievability of \eqref{eq:RSV}. The
proposed scheme, combines superposition coding, dirty paper
coding, and time-sharing, and exploits a representation of the
interferences in the form
\begin{equation}
\begin{aligned}
S_1^n &= A^n + D^n \\
S_2^n &= A^n - D^n,
\end{aligned}
\label{eq:int-synth}
\end{equation}
where
\begin{equation}
\begin{aligned}
A^n &= (S_1^n + S_2^n) / 2 \\
D^n &= (S_1^n - S_2^n) / 2.
\end{aligned}
\label{eq:int-anal}
\end{equation}

We list the main steps for codebook generation, encoding and
decoding. The probability of error analysis will be omitted as it
is based on standard typicality arguments. See
e.g.~\cite{CoverBook}.

{\bf Codebook Generation:} The idea is to generate three
codebooks. There is one common codebook which both the users share
and two private codebooks which are intended for the corresponding
user. More specifically we follow the following steps:
\begin{enumerate}
\item Decompose the message $W$ into two submessages $W_A$ and
$W_D$ and divide the power $P$ into two powers $P_A$ and $P_D$ so
that $P=P_A+P_D$. Message $W_A$ will be decoded by both the
receivers while message $W_D$ will be decoded by only one receiver
at a time. We will transmit it twice so that both the receivers
can decode (see encoding and decoding rules below for a further
description).

\item Generate a codebook $\cC_A$ for $W_A$ where the codewords
$U_A^n$ are sampled from i.i.d.\ a Gaussian distribution $U_A =
X_A + \alpha_A A$. Here $X_A$ is Gaussian $\cN(0,P_A)$,
independent of $A$ and $\alpha_A = P_A/(P+Q/2+1)$. A total of
$2^{n I(U_A;Y_i)}$ codewords are thus generated and randomly
partitioned into $2^{n I(U_A;A)}$ bins. The rate of this codebook,
$I(U_A;Y_i)-I(U_A;A)$ can be shown to be~\footnote{Using a
symmetry argument or otherwise, note that $I(U_A;Y_1)=I(U_A;Y_2)$,
so we use the generic term $I(U_A;Y_i)$ to denote either of
these.}:
  \begin{equation}
    R_A = \frac{1}{2} \log\left(1+\frac{P_A}{P_D+Q/2+1}\right).
    \label{eq:RS}
  \end{equation}

\item Generate two codebooks $\cC_D^{(1)}$ and $\cC_D^{(2)}$  for
$W_D$ for the two receivers as follows. For $\cC_D^{(1)}$, the
codewords $U_D^n$ are sampled from a i.i.d.\  Gaussian
distribution $U_D = X_D + \alpha_D ( (1-\alpha_A)A + D)$, where
$X_D$ is Gaussian $\cN(0,P_D)$, independent of $A$ and $D$ and
$\alpha_D = P_D/(P_D+1)$. Generate $2^{nI(U_D;Y_1,U_A)}$ such
codewords and partition them into $2^{n I(U_D;A,D)}$ bins. Follow
analogous construction for codebook $\cC_D^{(2)}$. The rate of
each codebook\footnote{Notice that the codebooks can be the same
for two users. For notational convenience while dealing with the
two users we keep the codebooks separate since a codeword typical
with $Y_1^n$ will not in general be typical with $Y_2^n$. See the
encoding rules below.} $I(U_D;Y_i,U_A)-I(U_D;A,D)$ can be shown to
be:
  \begin{equation}
    R_D = \frac{1}{2} \log(1+P_D).
    \label{eq:RV}
  \end{equation}
\end{enumerate}

{\bf Encoding:} We transmit a superposition of two sequences
corresponding to $W_A$ and $W_D$ as follows:
\begin{enumerate}
\item To encode a message $W_A$, find a codeword $U_A^n$ in the
bin of $W_A$, such that $X_A^n = U_A^n - \alpha_A A^n$ satisfies a
power constraint of $P_A$. By construction, such a codeword exists
with high probability.

\item To encode $W_D$, we decide whether to send it to user 1 or
2. The users are served alternately. When we decide to send it to
user 1, we select a codeword $U_D^n$ in the bin of codebook
$\cC_D^{(1)}$ corresponding to message $W_D$ such that  $X_D^n
  = U_D^n - \alpha_D\{(1-\alpha_A)A^n + D^n\}$ satisfies a power
  constraint of $P_D$.  When we decide to transmit to user 2, we
  select a codeword $U_D^n$ in the bin of codebook $\cC_D^{(2)}$
  corresponding to message $W_D$ such that $X_D^n
  = U_D^n - \alpha_D\{(1-\alpha_A)A^n - D^n\}$ satisfies the power
  constraint of $P_D$. Since there are $2^{nI(U_D;A,D)}$ codewords
  in each bin, such a codeword exists with high probability.

\item Send the superposition $X^n=X_A^n+X_D^n$, which has power
$P$, over the channel.

\end{enumerate}

{\bf Decoding:}
The decoding exploits successive cancellation
(stripping) and proceeds as follows:
\begin{enumerate}

\item Decode $U_A^n$ from $Y_1^n$ or $Y_2^n$ treating $X_D^n$
  as part of the noise.  The received signals are of the form
  \begin{align*}
     Y_1^n &= X_A^n + A^n + ( D^n + Z_1^n + X_D^n)\\
           &= U_A^n + (1-\alpha_A)A^n + ( D^n + Z_1^n + X_D^n),\\
     Y_2^n &= X_A^n + A^n + (-D^n + Z_2^n + X_D^n) \\
           &= U_A^n + (1-\alpha_A)A^n + ( -D^n + Z_2^n + X_D^n).
  \end{align*}
Since $D^n + Z_i^n + X_D^n$ is an i.i.d.\ Gaussian
$\cN(0,P_D+Q/2+1)$ sequence, independent of  $A^n$, our choice of
rate $R_A$ in \eqref{eq:RS} ensures that the resulting $\hat{W}_A$
equals $W_A$ with high probability at both the receivers.

\item Subtract the decoded $U_A^n$ from each of $Y_1^n$ and
$Y_2^n$, so  that the residual signals $\tilde{Y}_i^n =
Y_i^n-U_A^n$ are of the form
    \begin{align}
     \tilde{Y}_1^n &= X_D^n + ((1-\alpha_A)A^n + D^n) + Z_1^n, \\
     \tilde{Y}_2^n &= X_D^n + ((1-\alpha_A)A^n - D^n) + Z_2^n.
    \end{align}
The rate $R_D$ in \eqref{eq:RV} ensures that $U_D^n$ can be
decoded from either $\tilde{Y}_1^n$ or $\tilde{Y}_2^n$ so that the
resulting $\hat{W}_D$ equals $W_D$ with high probability at the
corresponding receiver.  Specifically, for the fraction of time
that the transmitter encodes $W_D$ for interference
$(1-\alpha_A)A^n+D^n$, user~1 can recover $W_D$, while for the
fraction of time that the transmitter encodes $W_D$ for
interference ${(1-\alpha_A)A^n-D^n}$, user~2 can recover $W_D$.

\end{enumerate}

From this coding strategy, we see that the average rate delivered
to each receiver is identical, i.e., $R_A + (1/2) R_D$. Maximizing
this rate over the choices of $P_A$ and $P_D$ subject to the
constraint $P=P_A+P_D$ optimizes the lower bound, whence
\eqref{eq:glb}.

\end{proof}
%

From \eqref{eq:glb-exp}, we obtain several useful insights. First,
note that in the high INR regime ($Q/2 \geq P+1$), our lower bound
reduces to time-sharing, while in the low INR regime ($Q/2 \leq
1$) it reduces to dirty paper coding with respect to $A^n$. In the
moderate interference regime, our bound shows that one can
generally achieve a gain over these two strategies by a
superposition coding approach that combines them.

\begin{figure}[tbp]
\begin{center}
\includegraphics[width=3.25in]{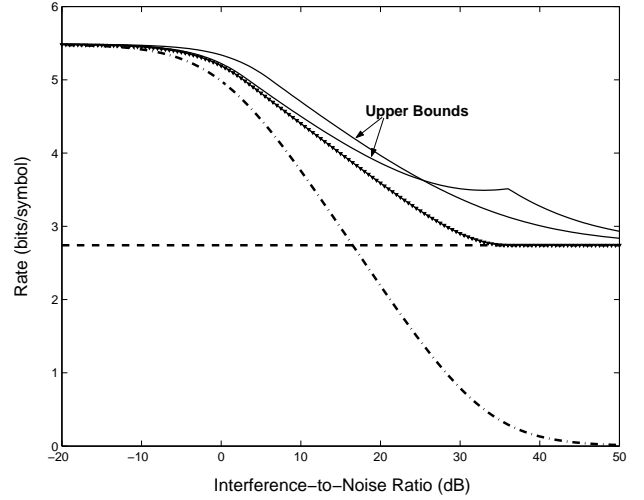}
\caption{Upper and lower bounds on the capacity of the two-user
Gaussian multicast channel, as a function of INR $Q$ for an SNR
$P=33$~dB.  The upper two curves depict the two upper bounds
from~\eqref{eq:gub_d} and~\eqref{eq:gub_s}. The marked line is the
achievable rate in~\eqref{eq:glb-exp}. The horizontal dashed line
indicates the performance of time-sharing, while the other dashed
curve indicates the performance of a strategy in which the side
information is treated by the transmitter as additional noise on
each link. \label{fig:gaussboundsinr}}
\end{center}
\end{figure}

The behavior of the bounds as a function of INR is depicted in
Fig.~\ref{fig:gaussboundsinr} for a fixed SNR of $P = 33$~dB. When
the INR is very small ($Q \ll 1$), Fig.~\ref{fig:gaussboundsinr}
reflects the rather obvious fact that the side information can be
ignored by the transmitter without sacrificing rate.  Similarly,
when the INR is large($Q \gg 1$), Fig.~\ref{fig:gaussboundsinr}
reflects that time-sharing  between the two users achieves the
capacity. More generally,

.
\begin{equation}
\lim_{Q\rightarrow\infty} C \leq \lim_{Q\rightarrow\infty}
R_+^\mrm{I} = \lim_{Q\rightarrow\infty}R_+^\mrm{II}=\frac{1}{4}
\log(1+P),
\end{equation}
which can be achieved by time-sharing between the two users and
doing Costa dirty paper coding for each user being served.
We note that this result settles the conjecture made in
\cite{SteinbergShamai05}.

\begin{figure}[tbp]
\begin{center}
\includegraphics[width=3.25in]{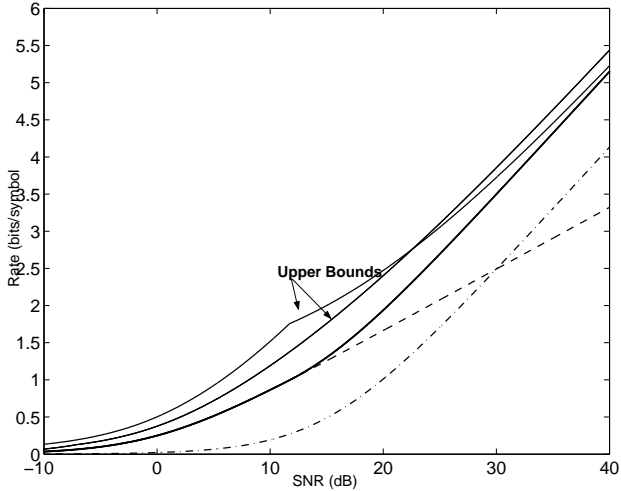}
\caption{Upper and lower bounds on the capacity of the two-user
Gaussian multicast channel, as a function of SNR $P$ for an INR $Q
= 15$~dB.  The upper two curves depict the two upper bounds
in~\eqref{eq:gub_d} and~\eqref{eq:gub_s}. The achievable rate
in~\eqref{eq:glb-exp} is also shown. The dashed curve indicates
the performance of time-sharing, while the dash-dotted curve
indicates the performance of a strategy in which the side
information is treated by the transmitter as additional noise on
each link. \label{fig:gaussboundssnr}}
\end{center}
\end{figure}

Perhaps more interestingly, our proposed achievable rate is
optimal in the limit of high SINR. The behavior of the bounds as a
function of SNR is depicted in Fig.~\ref{fig:gaussboundssnr} for a
fixed INR of $Q = 15$~dB. We note that the expression for
$R_+^\mrm{II}$ coincides with $R_-$ in this limit.  Note that the
base-line schemes do not achieve a rate particularly close to
capacity, but the superposition dirty paper coding strategy
corresponding to our lower bound does. More generally, we can show
that:
\begin{align}
\lim_{P\rightarrow \infty} (C-R_-) &\le \lim_{P\rightarrow \infty}
(R_+^\mrm{II}-R_-) =0 \label{eq:fixedQhighP}
\end{align}

To verify \eqref{eq:fixedQhighP} for $Q\ge2$,  since $P\rightarrow
\infty$, the middle case of the lower bound \eqref{eq:glb-exp}
applies which we can alternately express in the form
\begin{equation}
R_- = \frac{1}{2}\log\left(\frac{P+Q/2+1}{\sqrt{2Q}}\right)
\label{eq:middle}
\end{equation}
Comparing \eqref{eq:middle} with the upper bound \eqref{eq:gub_d}
we have
\begin{equation}
\begin{aligned}
R_+^\mrm{II}-R_- &=
\frac{1}{2}\log\frac{P+Q+1+2\sqrt{PQ}}{\sqrt{2Q}}\\
 &\quad- \frac{1}{2}\log\frac{P+Q/2+1}{\sqrt{2Q}},
\end{aligned}
\end{equation}
which in the limit $P\rightarrow\infty$ gives
\eqref{eq:fixedQhighP}. The case $Q\le 2$, can be similarly
verified. We summarize the optimality properties in the following
corollary.
\begin{corol}
For the Gaussian multicast channel in Figure~\ref{fig:gaussian},
the proposed achievable rate in
Theorem~\ref{thm:gauss_lower_bounds} is optimal in the limit of
high SINR ($P\rightarrow\infty, Q$ is fixed). For $Q>2$ it can be
expressed as $C(P) =
\frac{1}{2}\log\left(\frac{P}{\sqrt{2Q}}\right) + o(1) $, where
$o(1)\rightarrow 0$ as $P \rightarrow \infty$. For $Q \le 2$ it
can be expressed as $C(P) =
\frac{1}{2}\log\left(\frac{P}{1+Q/2}\right)+o(1)$. Finally, for
the case of fixed $P$ and $Q\rightarrow\infty$, time-sharing
between the two users is optimal and the capacity can be expressed
as $C(P) = \frac{1}{4}\log(1+P) + o(1)$, where $o(1)\rightarrow 0$
as $Q\rightarrow \infty$.
\end{corol}

Finally, we show in Appendix~\ref{app:c-UniversalGap} that a
universal constant that bounds the difference between our upper
and lower bounds is given by:
\begin{equation}
\sup_{P,Q} R_+^\mrm{II}-R_- =
\frac{1}{2}\log\left(\frac{3}{2}+\sqrt{2}\right) =
0.7716\label{eq:universalBound}
\end{equation}

We conclude this section with a few additional observations.

\subsubsection*{Some Further Remarks}

\begin{enumerate}

 \item {\emph{Extension to K
receivers}}: Our upper bounding technique for $R^\mrm{II}_+$
in~\eqref{eq:gub_d} can be extended to the case of $K$ receivers
each with independent interference. We show in
Appendix~\ref{app:c-Kuser} that the following upper bound holds
for the case of $K$ receivers:
\begin{equation}
\begin{aligned}
R^K_+ &\le \frac{1}{2}\log(P+Q+1+2\sqrt{PQ}) -\frac{K-1}{2K}\log Q \\
&\quad- \frac{1}{2K}\log K -
\left[\frac{1}{2K}\log\left(\frac{Q}{K(P+1)}\right)\right]^+
\end{aligned}
\label{eq:gauss_K_user_UB}
\end{equation}
By taking the limit $Q\rightarrow\infty$
in~\eqref{eq:gauss_K_user_UB}, it can be sown that time-sharing is
optimal for any number of users in the high INR limit.

\item  {\emph{Correlation between noise sequences}}: The upper
bound in Theorem~\ref{thm:gaussbounds} is valid even when the
noises $Z_1^n$ and $Z_2^n$ are not independent. The argument is
analogous to that for the standard broadcast channel (e.g.
\cite[Ch. 14]{CoverBook}). We exploit this observation to derive
the upper bound expressions. Furthermore analogous to the result
in~\cite{Cohen02}, even if the noise is not Gaussian our lower
bound in~\eqref{eq:glb} is achievable when the decoder treats the
noise as Gaussian.

\item {\emph{Feedback does not help much}}. As discussed in
Appendix~\ref{app:c-Feedback Argument} and~\ref{app:d-Feedback
Argument}, the expressions for $R_+^\mrm{I}$ and $R_+^\mrm{II}$
in~\eqref{eq:gub_s_opt} and~\eqref{eq:gub_d_opt} continue to hold
in the presence of perfect causal feedback, provided we do not
optimize over the parameter $\rho$, but set it to equal the actual
correlation between the noise terms.


\item The capacity-achieving strategy for the binary channel does not
  extend immediately to the Gaussian channel.   While one might
  speculate that an adaptation of the achievability approach in
  Theorem~\ref{thm:2usrbinary} for the Gaussian channel would improve
  on the lower bound \eqref{eq:glb} in Theorem~\ref{thm:gauss_lower_bounds},
  the obvious generalizations do not.  In particular, strategies which
  precancel the interference in part of the codeword for each user
  achieved lower rates than our superposition dirty paper
  coding; for a further discussion see \cite{khisti_thesis}.

\end{enumerate}

\subsection{Correlated Interferences and Robust Dirty Paper Coding}
\label{sec:rdpc}

Consider the a memoryless Gaussian point-to-point channel model with output
\begin{equation}
Y^n = X^n + { S}^n + Z^n, \label{eq:dpc}
\end{equation}
where $X^n$ is the channel input subject to power constraint $P$,
${S}^n$ is a white Gaussian interference sequence of power $Q$ not
known to decoder, and $Z^n$ is a white Gaussian noise sequence of
unit power.  When the interference ${S}^n$ is perfectly known to
the encoder, Costa's dirty paper coding is capacity achieving.
However, in many applications, only imperfect knowledge of ${
S}^n$ is available to the encoder.  One special case is the case
of \emph{causal} knowledge considered by Shannon. Another is the
case of \emph{noisy} noncausal knowledge. For these kinds of
generalizations, there is interest in understanding the capacity
of such channels and the structure of the associated
capacity-achieving codes, which we refer to as \emph{robust} dirty
paper codes.

It is often natural to analyze such problems via their equivalent
Gaussian multicast model.  As an illustration, suppose that the
interference in \eqref{eq:dpc} is of the form ${S}^n = \beta
S_0^n$ where $S_0^n \thicksim \cN(0,Q\bI)$ is known to the encoder
but $\beta$ is not. Then if $\beta$ is from a finite alphabet (or
can be approximated as being so), i.e.,
$\beta\in\{\beta_1,\beta_2,\dots,\beta_K\}$, the problem is
equivalent to a Gaussian multicast problem with $K$ users where
the interference for the $k$\/th user is $\beta_k S_0^n$.

From this example it is apparent that for at least some
applications, there is a need to accommodate \emph{correlated}
interferences in the Gaussian multicast model. In what follows we
focus on that case where there are two receivers i.e. $\beta \in
\{\beta_1,\beta_2\}$. Extensions to the case of more than two
receivers are possible, but will not be explored.

We first provide a general upper bound for the case of correlated,
jointly Gaussian interference sequences and then specialize it to
the case of scaled interferences. The general upper bound might be
of independent interest and is derived in
Appendix~\ref{appe-sec:corrInterf}.

\begin{thm}
Consider a two receiver channel model $Y_i^n = X^n + S_i^n + Z^n$
for $i=1,2$ when $Z^n$ is i.i.d.\ $\cN(0,1)$ noise, $S_1^n$ and
$S_2^n$ are i.i.d.\ jointly Gaussian with marginal distributions
$\cN(0,Q_1)$ and $\cN(0,Q_2)$ respectively and suppose that the
distribution of $S_1-S_2$ is $\cN(0,Q_d)$. An upper bound on the
common message rate for this channel under a power constraint $P$
at the transmitter is given by:
\begin{equation}
\ R_+^\mrm{C} = \sum_{i=1}^2 \frac{1}{4}\log(P+Q_i+1+2\sqrt{PQ_i})
- T(Q_d) \label{eq:CorrelatedInter_UB}
\end{equation}
where
\begin{equation}
T(Q_d)=
\begin{cases}
\frac{1}{4}\log(Q_d), & Q_d > 4 \\
\frac{1}{2}\log\left(1+\frac{Q_d}{4}\right), & Q_d \le 4
\end{cases}
\label{eq:Tq_expr}
\end{equation}
 \label{thm:CorrelatedInterf}
\end{thm}

We note that the upper bound is of most interest in the high
signal-to-interference-plus-noise limit i.e. when we fix
$Q_1$,$Q_2$ and take $P\rightarrow\infty$. In this limit we have
the following:
\begin{corol}
In the high SINR limit ($Q_1,Q_2$ fixed, $P\rightarrow\infty$),
the upper bound on the case of correlated interferences in
Theorem~\ref{thm:CorrelatedInterf} can be written as
\begin{equation}
R_+^\mrm{C} = \frac{1}{2}\log(P) - T(Q_d)+ o(1),
\label{eq:CorrelatedInter_UB_HIGH_SINR}
\end{equation}
where the term $o(1)$ approaches 0 as $P \rightarrow \infty$ and
$Q_1,Q_2$ fixed and $T(Q_d)$ is given in~\eqref{eq:Tq_expr}.
\end{corol}

To establish an achievable rate, we will consider a modification
to our lower bound in Theorem~\ref{thm:gauss_lower_bounds} which
considers the case of independent interferences. To deal with the
case of correlated interferences, we will require that the encoder
and decoders have access to a common source of randomness which
will be used as a dither sequence.

Consider a superposition dirty paper coding strategy analogous to
that in the proof of the lower bound in
Theorem~\ref{thm:gauss_lower_bounds}, whereby we decompose the
interferences according to \eqref{eq:int-synth}. In this case, we
have that \eqref{eq:int-anal} specializes to
\begin{equation}
\begin{aligned}
A^n &= \beta_A\,S_0^n \\
D^n &= \beta_D\,S_0^n,
\end{aligned}
\label{eq:int-anal-beta}
\end{equation}
where
\begin{equation}
\begin{aligned}
\beta_A &= (\beta_1 + \beta_2) / 2 \\
\beta_D &= (\beta_1 - \beta_2) / 2.
\end{aligned}
\label{eq:int-synth-beta}
\end{equation}

When we turn to implement the encoding step in the proof of the
lower bound of Theorem~\ref{thm:gauss_lower_bounds}, in which
$A^n$ is treated as interference and $D^n$ as noise, the results
of \cite{costa1983} cannot be directly applied since the
interferences $A^n$ and $D^n$ in \eqref{eq:int-anal-beta} are
correlated. On the other hand, if we assume that the encoder and
decoder(s) have access to a source of common randomness in the
form of a dither sequence, we can use the lattice coding strategy
in~\cite{ErezShamaiZamir01}. In this scheme, the transmitted
sequence is statistically independent of the interference and
noise sequences. It can be easily shown that for such schemes,
correlation between the interference and noise sequences does not
change the achievable rate relative to the case when the noise and
interference sequences are independent~\footnote{In fact, the
result in~\cite{ErezShamaiZamir01} holds for an arbitrary
interference sequence.}. With this scheme, we obtain the following
lower bound.
\begin{thm}
An achievable rate for our example multicast channel with
correlated interferences and common randomness at the encoder and
decoders is given by:
\begin{subequations}
\begin{equation}
C^\beta(P) \ge \max_{\{(P_A,P_D): P_A\ge0, P_D\ge0, P_A+P_D\le
P\}} R^\beta(P_A,P_D),
\end{equation}
where
\begin{equation}
\begin{aligned}
R^\beta(P_A,P_D)  &=
\frac{1}{2}\log\left(1 + \frac{P_A}{1+Q_d/4+P_D}\right) \\
 &\quad + \frac{1}{4}\log\left(1 + P_D\right).
\end{aligned}
\end{equation}%
\label{eq:GaussianCorr}%
\end{subequations}
where $Q_d \defeq (\beta_1-\beta_2)^2Q$ is the variance of
$S_1-S_2$. \label{thm:CorrelatedInterf_achievable}
\end{thm}
Optimizing over $P_A$ and $P_D$, gives the following achievable
rate:
\begin{equation}
R_-^\beta(P)=
\begin{cases}
\frac{1}{2}\log\left(1+\frac{P}{1+ Q_d/4}\right), & Q_d <4\\
\frac{1}{2}\log\left(\frac{P+1+Q_d/4}{\sqrt{Q_d}}\right), & 4\le
Q_d \le
4(P+1)\\
\frac{1}{4}\log(1 + P), &  Q_d \ge 4P+4
\end{cases}
\label{eq:Correlated_Interf_LowerBound}
\end{equation}

We note that in the limit of high SINR, our expression for
$R_-^\beta$ in~\eqref{eq:Correlated_Interf_LowerBound} is given by
$R_-^\beta= \frac{1}{2}\log(P) - T(Q_d) + o(1)$, where $T(Q_d)$ is
given as in~\eqref{eq:Tq_expr}. This coincides with the upper
bound in~\eqref{eq:CorrelatedInter_UB_HIGH_SINR} and thus
establishes the optimality of our scheme in the high SINR limit.
\begin{corol}
The proposed achievable rate in
Theorem~\ref{thm:CorrelatedInterf_achievable} is optimal in the
limit of high SINR (fixed $Q_1,Q_2$, $P\rightarrow\infty$) i.e.\
$\lim_{P\rightarrow\infty}C^\beta(P)-R_-^\beta(P) =0.$
\end{corol}

\section{Concluding Remarks}
\label{sec-Conclusion}

We introduced the multicast channel model and analyzed the special
cases of binary and Gaussian channels with additive interference.
Our main observation in this work is that unlike the single user
case, the lack of side information at the receiver strongly limits
capacity. We show that in both the binary and Gaussian cases if
the interfering sequences are independent, time-sharing is optimal
in the limit of large interference. Also certain achievable rates
and their optimality properties have been discussed. The capacity
has been established for the two user noiseless binary case and
for the Gaussian case in the high
signal-to-interference-plus-noise ratio limit. Somewhat
surprisingly, the optimal schemes are very different for the two
cases.

It may be possible to extend the upper bounding techniques in this
paper to more general channel models and perhaps also sharpen the
results for the Gaussian and binary cases. We emphasize however
that the proposed bounds indicate an important engineering insight
that there is a significant loss in dealing with more than one
interference sequence at the transmitter, even when they are
correlated. An interesting direction of future work would be to
investigate the connections of this result with a recent result on
MIMO broadcast channel with imperfect channel state information at
the transmitter~\cite{LapidothShamaiWigger05}, where again it was
shown that lack of perfect CSI strongly limits the broadcast
channel capacity.

\appendices

\section{Proof of the Converse in Theorem~\ref{thm:2usrbinary}}
\label{app:a}

We have to show that for any sequence of $(2^{nR},n)$ codes with
$P_e^n\rightarrow 0$, we must have $R \leq C$, where $C$ is
defined in \eqref{eq:2usrCapacity}.

Since each receiver is able to decode the message we have from
Fano's inequality

\begin{equation}
H(W|Y_k^N) \leq n\eps_n, \hspace{2em} \text{ for } k=1,2,
\label{eq:bin-conv-fano}
\end{equation}
where $\eps_n$ is a sequence that approaches $0$ as $n\rightarrow
\infty$. We can use Fano's inequality to bound the rate as
\begin{align}
nR  &= H(W) \notag\\
&= H(W|Y_1^n) + I(W;Y_1^n) \notag\\
&\leq n\eps_n + H(Y_1^n) - H(Y_1^n|W) \label{eq:bin-conv-1}\\
&\leq n\eps_n + \sum_{j=1}^n H(Y_{1j}) - H(Y_1^n|W) \label{eq:bin-conv-2}\\
&\leq n\eps_n + n - H(Y_1^n|W), \label{eq:bin-conv-3}
\end{align}
where \eqref{eq:bin-conv-1} follows by using the Fano inequality
\eqref{eq:bin-conv-fano}, \eqref{eq:bin-conv-2} follows from the chain
rule and the fact that conditioning reduces the entropy, and
\eqref{eq:bin-conv-3} follows from the fact that each $Y_{1j}$ is
binary valued.  We can similarly bound the rate on the second user's
channel as
\begin{equation}
nR \leq n\eps_n + n - H(Y_2^n|W).
\label{eq:bin-conv-4}
\end{equation}
Combining \eqref{eq:bin-conv-3} and \eqref{eq:bin-conv-4}, we
obtain
\begin{align}
  nR  &\leq n - \max\{H(Y_1^n|W), H(Y_2^n|W)\} + n\epsilon_n \notag\\
        &\leq n - \frac{1}{2}\{H(Y_1^n|W)+ H(Y_2^n|W)\} + n\epsilon_n  \notag \\
        &\leq n - \frac{1}{2}H(Y_1^n,Y_2^n|W) + n\epsilon_n      \label{eq:BinProof1}        \\
        &\leq n - \frac{1}{2}H(Y_1^n \oplus Y_2^n|W) + n\epsilon_n  \label{eq:BinProof2}\\
        &= n - \frac{1}{2}H(S_1^n \oplus S_2^n) + n\epsilon_n  \label{eq:BinProof3} \\
        &= n\left(1-\frac{1}{2}H(S_1\oplus S_2) +
        \epsilon_n\right), \label{eq:BinProof4}
\end{align}
where \eqref{eq:BinProof1} follows from the fact that conditioning
reduces entropy, \eqref{eq:BinProof2} follows from the fact
that $Y_1^n \oplus Y_2^n$ is a deterministic function of
$(Y_1^n,Y_2^n)$, \eqref{eq:BinProof3} follows from the fact
that $Y_1 \oplus Y_2 = S_1 \oplus S_2$, and \eqref{eq:BinProof4}
follows from the fact that both $S_1$ and $S_2$ are i.i.d.\ so the
joint entropy of the sequence $S_1^n \oplus S_2^n$ is the sum of
the individual terms.

\section{Proof of upper bound \protect\eqref{eq:bub} in Theorem~\ref{thm:kuserBinary}}
\label{app:b}

The upper bound mirrors the converse for two-user case.  In
particular, following the same steps as in the two-user case to
derive \eqref{eq:BinProof1}, we have that any achievable rate
satisfies
\begin{equation}
nR \leq n - \frac{1}{K}H(Y_1^n,Y_2^n,\ldots, Y_K^n|W) + n\epsilon_n.
\label{eq:abc}
\end{equation}
Proceeding from \eqref{eq:abc} we obtain
\begin{align}
  & nR -n\epsilon_n  \nonumber\\
  &\leq n- \frac{1}{K}H(Y_1^n,Y_2^n,\ldots, Y_K^n|W) \nonumber\\
 &= n - \frac{1}{K}H(Y_1^n,Y_1^n\oplus Y_2^n,\ldots,Y_1^n\oplus Y_K^n|W) \label{eq:bin-k-conv-1}\\
&= n - \frac{1}{K}H(X^n\oplus S_1^n,S_1^n\oplus
S_2^n,\ldots,S_1^n\oplus S_K^n|W) \notag\\
&= n -\frac{1}{K}H(S_1^n\oplus S_2^n,\ldots,S_1^n\oplus S_K^n|W) \nonumber\\
&\quad\quad - \frac{1}{K}H(X^n\oplus S_1^n|S_1^n\oplus S_2^n,\ldots S_1^n\oplus S_K^n,W)\notag\\
&=n - \frac{n}{K}H(S_1\oplus S_2,\ldots, S_1\oplus
S_K) \nonumber\\
&\quad\quad - \frac{1}{K}H(X^n\oplus S_1^n|S_1^n\oplus
S_2^n,\ldots S_1^n\oplus
S_K^n,W) \label{eq:bin-k-conv-2}\\
&\leq n - \frac{n}{K}H(S_1\oplus S_2,\ldots, S_1\oplus S_K)\notag,
\end{align}
where \eqref{eq:bin-k-conv-1} follows from the fact that the
mapping $(Y_1^n,Y_2^n,\ldots Y_K^n) \rightarrow (Y_1^n,Y_1^n\oplus
Y_2^n,\ldots,Y_1^n\oplus Y_2^n)$ is invertible, and
\eqref{eq:bin-k-conv-2} follows from the fact that
$S_1^n,S_2^n,\ldots S_K^n$ are all i.i.d.\ and independent of $W$.

\section{Proof of Upper Bound \protect\eqref{eq:gub_d} in Theorem~\protect\ref{thm:gaussbounds}}
\label{app:c}

We now derive~\eqref{eq:gub_d} for $R_+^\mrm{II}$. We first note
that the capacity of the channel only depends on the marginal
distributions $p(Y_1^n|X^n,S_1^n,S_2^n)$ and
$p(Y_2^n|X^n,,S_1^n,S_2^n)$ and not on the joint distribution
$p(Y_1^n,Y_2^n|X^n,S_1^n,S_2^n)$. Allowing correlation between the
noise $Z_1$ and $Z_2$ does not change capacity. Specifically, we
have
\begin{lemma}
\label{lem:CorrelDoesNotMAtter} Let $P_e^n$ be the probability of
decoding error in \eqref{eq:errprob}. If $P_e^n$ is bounded away
from zero for a certain correlation between $Z_1$ and $Z_2$ above
then it is bounded away from zero for \emph{any} other correlation
between $Z_1$ and $Z_2$.
\end{lemma}

\begin{proof}
The argument is essentially the same as given in \cite[Ch 14, Page
454]{CoverBook}. We repeat it here for completeness. Let
$P_e^{1,n}$ and $P_e^{2,n}$ denote the error probabilities in
decoding at receiver 1 and 2 respectively. We have
\begin{align*}
P_e^{1,n} &= \Pr\left(g_1(Y_1^n)\neq W\right) \\
P_e^{2,n} &= \Pr\left(g_2(Y_2^n) \neq W\right) \\
P_e^n &= \Pr\left(\bigcup_{k=1,2}\{ g_k(Y_k^n) \neq W\}\right)
\end{align*}

Next, note that
\begin{equation}
 \max\{P_e^{1,n},P_e^{2,n}\} \leq P_e^n \leq P_e^{1,n} + P_e^{2,n},
 \label{eq:marginalBounds}
\end{equation}
where the left inequality in \eqref{eq:marginalBounds} follows
from the fact that by definition $P_e^n \geq P_e^{k,n}$ for
$k=1,2$, and the right inequality follows from the union bound. In
turn, note that both $P_e^{1,n}$ and $P_e^{2,n}$ do not depend on
the correlation between $Z_1$ and $Z_2$.  Accordingly, both the
left and right hand terms in \eqref{eq:marginalBounds} do not
depend on the correlation between $Z_1$ and $Z_2$. In particular
if $P_e^n$ is bounded away from $0$ for some correlation between
$Z_1$ and $Z_2$, then necessarily one of $P_e^{1,n}$ and
$P_e^{2,n}$ is bounded away from zero. Thus the probability of
error is bounded away from zero for all possible correlations.
\end{proof}

In the rest of the section we will fix $E[Z_1 Z_2] = \rho$ and
derive an upper bound. Thereafter, we will optimize over $\rho$,
to tighten the upper bound.  We will need the following additional
properties of $Z_1$ and $Z_2$, which are readily computed.
\begin{lemma}
\label{lem:Z+Z-} Let $Z_1$ and $Z_2$ be standard normal, jointly
Gaussian random variables with correlation $\rho$.  Define $Z_-
\defeq (Z_1-Z_2)/\sqrt{2}$ and $Z_+ \defeq (Z_1 + Z_2)/\sqrt{2}$.
Then $Z_+$ and $Z_-$ are independent zero-mean Gaussian random
variables with variances $1+\rho$ and $1-\rho$, respectively.
\end{lemma}

To obtain our upper bound we show that a sequence of $(2^{nR},n)$
codes that can be decoded by both the receivers with
$P_e^n\rightarrow 0$ must satisfy $R \leq R_+^\mrm{II}$ in
\eqref{eq:gub_d_opt}. Note that our power constraint is of the
form $E[X_i^2] \le P_i$ with $\sum_{i=1}^n P_i \le nP$.

Suppose $R_1$ and $R_2$ denote the rates at which the two
receivers can reliably decode the common message. The rate of the
common message must satisfy $R \leq \min(R_1,R_2)$.

From Fano's inequality, we have that for some sequence $\eps_n$,
which approaches 0 as $n\rightarrow \infty$,
\begin{equation}
H(W|Y_k^n) \leq n\eps_n, \hspace{2em} \text{ for } k=1,2.
\label{eq:gauss-conv-fano}
\end{equation}


We first upper bound $R_1$ as
\begin{align}
nR_1 &< I(W;Y_1^n) + n\epsilon_n \nonumber\\
&= h(Y_1^n) - h(Y_1^n|W) + n\epsilon_n \nonumber\\
&\leq \sum_{i=1}^n h(Y_i) - h(Y_1^n|W) + n\epsilon_n \label{eq:g-conv-r1-1}\\
 &\leq \sum_{i=1}^n\frac{1}{2}\log 2\pi e(P_i+1+ Q+ 2\sqrt{P_i Q})-h(Y_1^n|W)+n\epsilon_n.\label{eq:g-conv-r1-11}\\
 &\leq \frac{n}{2}\log 2\pi e(P+1+Q + 2\sqrt{PQ})-h(Y_1^n|W)+n\epsilon_n,\label{eq:g-conv-r1-F}
\end{align}
where \eqref{eq:g-conv-r1-1} follows from the chain rule and the
fact that conditioning reduces entropy, and
\eqref{eq:g-conv-r1-11} follows from the fact that each $Y_i$ has
a variance no larger than $P_i+1+Q + 2\sqrt{P_iQ}$ and its
differential entropy can be upper bounded by that of a Gaussian
RV.  Finally, \eqref{eq:g-conv-r1-F} is a consequence of Jensen's
inequality.

Similarly applying the above chain of inequalities on User 2, we
have
\begin{equation}
nR_2 \leq \frac{n}{2} \log 2\pi e(P+1+Q +
2\sqrt{PQ})-h(Y_2^n|W)+n\epsilon_n.\label{eq:g-conv-r2-F}
\end{equation}

Now we can find an upper bound on the common information rate
using~\eqref{eq:g-conv-r1-F} and~\eqref{eq:g-conv-r2-F}:
\begin{align}
nR &= n\min(R_1,R_2) \le \frac{n}{2}(R_1 + R_2) \nonumber\\
&\le \frac{n}{2} \log 2\pi e(P+1+Q + 2\sqrt{PQ}) - \frac{1}{2}h(Y_1^n|W) \nonumber\\
&\quad-
\frac{1}{2}h(Y_2^n|W) + n\eps_n \nonumber\\
&\le \frac{n}{2} \log 2\pi e(P+1+Q + 2\sqrt{PQ})-
\frac{1}{2}h(Y_1^n,Y_2^n|W) + n\eps_n \label{eq:commonRateTrick}
\end{align}
where the last inequality~\eqref{eq:commonRateTrick} follows from
the fact that conditioning reduces the differential entropy.

We now need to lower bound $h(Y_1^n,Y_2^n|W)$. In what follows we
will also use the notation $S_+^n = \frac{S_1^n +
S_2^n}{\sqrt{2}}$ and $S_-^n = \frac{S_1^n-S_2^n}{\sqrt{2}}$. Note
that $S_+$ and $S_-$ are mutually independent, Gaussian
$\cN(0,Q)$.
\begin{align}
&\quad h(Y_1^n,Y_2^n|W)\nonumber\\
&=h\left(\frac{Y_1^n-Y_2^n}{\sqrt{2}},\frac{Y_1^n+Y_2^n}{\sqrt{2}}\Biggm|W\right)\label{eq:entropyInvariance}\\
&= h(S_-^n + Z_-^n, \sqrt{2}X^n + S_+^n + Z_+^n | W)
\label{eq:substitution}\\
&= h(S_-^n + Z_-^n|W) + h(\sqrt{2}X^n + S_+^n + Z_+^n | W,S_-^n +
Z_-^n)\label{eq:chainRule}\\
&=h(S_-^n + Z_-^n)+ I(S_+^n; \sqrt{2}X^n + S_+^n +
Z_+^n|W,S_-^n+Z_-^n) \nonumber\\
&\quad + h(\sqrt{2}X^n + S_+^n +
Z_+^n|W,S_-^n+Z_-^n,S_+^n)\label{eq:chainRule2}\\
&\ge h(S_-^n + Z_-^n)+ I(S_+^n; \sqrt{2}X^n + S_+^n +
Z_+^n|W,S_-^n+Z_-^n) \nonumber\\
&\quad + h(\sqrt{2}X^n + S_+^n +
Z_+^n|W,S_-^n+Z_-^n,S_+^n,X^n)\label{eq:conditioning}\\
&= h(S_-^n + Z_-^n) + I(S_+^n; \sqrt{2}X^n + S_+^n +
Z_+^n|W,S_-^n+Z_-^n) \nonumber\\
&\quad+ h(Z_+^n) \label{eq:Z+isIndep}
\end{align}
The above steps are justified as follows.
In~\eqref{eq:entropyInvariance} we have used the fact that the
differential entropy is invariant to a transformation of unit
determinant. We substitute for $Y_1$ and $Y_2$
in~\eqref{eq:substitution}. ~\eqref{eq:chainRule} follows from the
chain rule. In ~\eqref{eq:chainRule2}, we first drop the
conditioning over $W$ in the first term, since $(S_-^n,Z_-^n)$ are
jointly independent of $W$ and expand the second term.
Finally~\eqref{eq:conditioning} follows from the fact that
conditioning on $X^n$ further reduces the differential entropy
while~\eqref{eq:Z+isIndep} is a consequence from $Z_+^n$ being
independent of $(X^n,S_+^n,S_-^n,Z_-^n,W)$.

Since $S_-^n,Z_+^n,Z_-^n$ are all i.i.d.\ Gaussian with powers
$Q$, $1+\rho$ and $1-\rho$ respectively, we have
from~\eqref{eq:Z+isIndep}
\begin{equation}
\begin{aligned}
h(Y_1^n,Y_2^n|W) &\ge I(S_+^n; \sqrt{2}X^n + S_+^n +
Z_+^n|W,S_-^n+Z_-^n) \\
 &\quad + \frac{n}{2}\log 2\pi e(Q+1-\rho) + \frac{n}{2}\log 2\pi e (1 + \rho)
\end{aligned}
\label{eq:hy1y2bound}
\end{equation}

It remains to lower bound the mutual information term
in~\eqref{eq:hy1y2bound}. We first note that since $S_+^n$ is
independent of  $(W,S_-^n,Z_-^n)$ one can drop the conditioning in
the mutual information expression.

\begin{lemma}
For each $n\ge1$ and for any distribution $p(X^n|S_-^n,S_+^n,W)$
such that $\sum_{i=1}^n E[X_i^2] \le nP$, The mutual information
term in~\eqref{eq:hy1y2bound} can be lower bounded as
\begin{equation}
\begin{aligned}
&\quad I(S_+^n; \sqrt{2}X^n + S_+^n + Z_+^n|W,S_-^n+Z_-^n) \\
&\ge I(S_+^n;\sqrt{2}X^n + S_+^n + Z_+^n) \ge
\left[\frac{n}{2}\log\left(\frac{Q}{2P + 1 + \rho}\right)\right]^+
\label{eq:MIConditioning}
\end{aligned}
\end{equation}
\label{eq:LemmaRD}
\end{lemma}

\begin{proof}
The left hand inequality follows immediately by expanding
$I(S_+^n; \sqrt{2}X^n + S_+^n + Z_+^n|W,S_-^n+Z_-^n)$  and using
the fact that $S_+^n$ is independent of $(S_-^n,Z_-^n,W)$.

The right-hand side is a consequence of the rate-distortion
theorem for i.i.d.\ Gaussian sources. Note that
$E[\sum_{i=1}^n(\sqrt{2}X_i + Z_{+i})^2]\le n(2P+1+\rho)$. Thus if
the right inequality were violated, for a certain distribution
$p(X^n|S_+^n)$, we could use it as a test channel in quantizing a
n-dimensional i.i.d.\ Gaussian source and do better than the rate
distortion bound. Alternately, note that
\begin{align}
&\quad I(S_+^n;\sqrt{2}X^n + S_+^n + Z_+^n) \nonumber\\
&= h(S_+^n) - h(S_+^n |
\sqrt{2}X^n + S_+^n + Z_+^n)\nonumber\\
&= h(S_+^n) - h(\sqrt{2}X^n + Z_+^n |\sqrt{2}X^n + S_+^n + Z_+^n)\label{eq:RDB1}\\
&\ge h(S_+^n) - h(\sqrt{2}X^n + Z_+^n ) \label{eq:RDB2}\\
&\ge h(S_+^n) - \sum_{i=1}^n h(\sqrt{2}X_i + Z_{+,i}) \label{eq:RDB3}\\
&\ge \frac{n}{2}\log  Q - \sum_{i=1}^n \frac{1}{2}\log (2P_i + 1 + \rho) \label{eq:RDB4}\\
&\ge \frac{n}{2}\log  Q -  \frac{n}{2}\log (2P+1 + \rho) \label{eq:RDB5}\\
&= \frac{n}{2}\left[\log\left(\frac{Q}{2P+1+\rho}\right)\right]^+
\end{align}
Here~\eqref{eq:RDB1} follows from the fact that $h(X|Y) =
h(Y-X|Y)$,~\eqref{eq:RDB2} from the fact that removing the
conditioning on $\sqrt{2}X^n + S_+^n + Z_+^n$ only increases the
differential entropy,~\eqref{eq:RDB3} follows from the chain
rule,~\eqref{eq:RDB4} follows from the fact that the differential
entropy with a fixed variance is maximized for a Gaussian
distribution and~\eqref{eq:RDB5} follows from Jensen's inequality.
This establishes~\eqref{eq:MIConditioning}.
\end{proof}

Finally, by
substituting,~\eqref{eq:MIConditioning},~\eqref{eq:hy1y2bound}
into~\eqref{eq:commonRateTrick}, we get
\begin{align}
R &\le
\frac{1}{2}\log\left(\frac{P+Q+1+2\sqrt{PQ}}{\sqrt{(Q+1-\rho)(1+\rho)}}\right)\nonumber\\
&\quad -
\left[\frac{1}{4}\log\left(\frac{Q}{2P+1+\rho}\right)\right]^+
+\eps_n \label{eq:R_+^{II}_Final}
\end{align}

Finally, since $\rho$ is a free parameter of choice, we can select
it to be the value that minimizes~\eqref{eq:R_+^{II}_Final} and
thus~\eqref{eq:gub_d_opt} follows. To obtain the tightest possible
bound we can optimize over the value of $\rho$. We
obtain~\eqref{eq:gub_d} by selecting the following choice for
$\rho$:
\begin{equation}
\rho^*(Q) =
\begin{cases}
Q/2 & \text{ if } Q \leq 2 \\
1   & \text{ if } Q > 2.
\end{cases}
\label{eq:opt-rho_2}
\end{equation}

\subsection{Gains from Feedback}
\label{app:c-Feedback Argument}  In the presence of feedback, the
transmitted symbol at time $i$ depends on the past output i.e.
$x_i = f(w,y_1^{i-1},y_2^{i-1},s^n)$. In this situation $Z_{+,i}$
is still independent of $(W,Z_{-}^n,S^n,X_1^i)$. This condition
suffices, for deriving the bounds
in~\eqref{eq:commonRateTrick},~\eqref{eq:hy1y2bound}
and~\eqref{eq:MIConditioning}. Lemma~\ref{lem:CorrelDoesNotMAtter}
does not hold however, since now the joint distribution between
noise sequences does matter in the probability of error. So while
the expression~\eqref{eq:R_+^{II}_Final} holds, one cannot
optimize over $\rho$, but must select the value to be the actual
correlation coefficient in the channel.

\subsection{Universal Gap between Upper and Lower Bounds}
\label{app:c-UniversalGap} In this section we
verify~\eqref{eq:universalBound}, the gap between upper and lower
bounds for all values of $P$ and $Q$. We consider three different
cases.

For $Q\le 2$, we have
\begin{align}
R_+^\mrm{II}-R_- &=
\frac{1}{2}\log\left(\frac{P+Q+1+2\sqrt{PQ}}{P+1+Q/2}\right)
\label{eq:boundGap}
\end{align}
It can be verified that the maximum for $P\ge0$ and $0\le Q \le 2$
occurs for $Q=2$ and $P = 1/4(9-\sqrt{17})$. The maximum value is
$1/2\log((5+\sqrt{17})/4) \approx 0.5947$.

For the case $2\le Q \le 2(P+1)$ the difference is also given
by~\eqref{eq:boundGap}. The supremum is attained when we set
$Q=2(P+1)$ and let $P\rightarrow \infty$. The supremum value is
$1/2\log((3+2\sqrt{2})/2) \approx 0.7716$.

Finally for the case $Q\ge 2(P+1)$, the difference between the
bounds is given by
\begin{align*}
R_+^\mrm{II}-R_- &=
\frac{1}{2}\log\left(\frac{P+Q+1+2\sqrt{PQ}}{Q}\right)
\end{align*}
The supremum is obtained by taking $Q =2(P+1)$ and letting
$P\rightarrow \infty$ and again equals $1/2\log((3+2\sqrt{2})/2)$.

\subsection{The case of K receivers}
\label{app:c-Kuser} We consider the case where there are $K$
receivers. To get an upper bound, we assume perfect correlation
between the noise sequences i.e. receiver $k=1,2,\ldots K$ gets
$Y_k^n = X^n + S_k^n + Z^n$, where the interferences $S_k^n$ are
mutually independent and i.i.d.\ $\cN(0,Q)$ and $Z^n$ is i.i.d.\
$\cN(0,1)$.

To upper bound the common rate for the case of $K$ receivers,
first note that the derivation that leads
to~\eqref{eq:commonRateTrick} can be straightforwardly generalized
to yield
\begin{equation}
\begin{aligned} nR &\le \frac{n}{2}\log 2\pi e(P+Q+1+2\sqrt{PQ}) \\
&\quad- \frac{1}{K}h(Y_1^n,Y_2^n,\ldots Y_K^n|W)+n\eps_n
\end{aligned}
\label{eq:commonRateTrick_Kuser}
\end{equation}
We now consider generalizing our derivation for
~\eqref{eq:hy1y2bound} to lower bound $h(Y_1^n,Y_2^n,\ldots
Y_K^n|W)$. Let us consider a set of $K$ orthogonal vectors $\bv_1,
\bv_2, \ldots \bv_K$, where $\bv_1 =
\frac{1}{\sqrt{K}}[1,1,\ldots,1]$ and $\bv_2,\ldots \bv_K$ are
arbitrarily chosen. Let $\bY^n = (Y_1^n,Y_2^n,\ldots,Y_K^n)$
denote the $K-$tuple of received sequences.
\begin{claim}
The component-wise inner product of $\bY^n$ with
$\bv_1,\ldots,\bv_K$ satisfies:
\begin{equation}
\begin{aligned}
\ip{\bY^n}{\bv_1} &= \sqrt{K}X^n+\sqrt{K}Z^n + T_1^n\\
\ip{\bY^n}{\bv_j} &= T_j^n \hspace{1em} \text{for } j=2,3,\ldots
K.
\end{aligned}
\label{eq:ComponentWiseIP}
\end{equation}
Where $T_1^n,T_2^n,\ldots T_K^n$ are mutually independent, i.i.d.\
Gaussian $\cN(0,Q)$ sequences. \label{claim:KuserGaussClaim}
\end{claim}
\begin{proof}
The expression for $\ip{\bY^n}{\bv_1}$ can be verified by direct
substitution. Here $T_1^n = \frac{1}{\sqrt{K}}(S_1^n+S_2^n+\ldots
+ S_K^n) $.  Since $\bv_j$ and $\bv_1$ are mutually orthogonal for
$j\ge 2$, we have $\sum_{i=1}^K v_{ji}=0$. Hence
$\ip{\bY^n}{\bv_j} = \sum_{i=1}^K v_{ji}S_i^n $. We denote $T_j^n
= \sum_{i=1}^K v_{ji}S_i^n$. Since the $S_j^n$ are mutually
independent and i.i.d.\ and $\bv_j$ are mutually orthogonal it
follows that $T_j^n$ are all mutually independent and i.i.d.\
$\cN(0,Q)$.
\end{proof}

We can now lower bound $h(Y_1^n,Y_2^n,\ldots Y_K^n|W)$ in a manner
analogous to the derivation in~\eqref{eq:hy1y2bound}.
\begin{align}
&\quad h(Y_1^n,Y_2^n,\ldots
Y_K^n|W)\nonumber\\
&=h(\ip{\bY_1^n}{\bv_1},\ip{\bY_2^n}{\bv_2},\ldots\ip{\bY_K^n}{\bv_K}|W)\label{eq:Kuser_Ortho}\\
&= h(\sqrt{K}X^n+\sqrt{K}Z^n+ T_1^n, T_2^n,\ldots T_K^n|W)\label{eq:Kuser_claim}\\
&= h(T_2^n) + \ldots + h(T_K^n) \nonumber\\
&\quad+ h(\sqrt{K}X^n+\sqrt{K}Z^n+
T_1^n | T_2^n,\ldots,T_K^n,W)\label{eq:Kuser_Independence}\\
&=\frac{n(K-1)}{2}\log 2\pi e Q  \nonumber\\
&\quad+ h(\sqrt{K}X^n+\sqrt{K}Z^n+
T_1^n|W,\{T_j^n\}_{j=2}^K) \label{eq:Kuser_IID}\\
&=\frac{n(K-1)}{2}\log 2\pi e Q  \nonumber \\
&\quad+ h(\sqrt{K}X^n+\sqrt{K}Z^n+
T_1^n|W,\{T_j^n\}_{j=1}^K) \nonumber\\
&\quad\quad+ I(T_1^n ;\sqrt{K}X^n+\sqrt{K}Z^n+ T_1^n|T_2^n\ldots
T_K^n,W) \label{eq:Kuser_MutInf_ID}\\
&\ge\frac{n(K-1)}{2}\log 2\pi e Q + \frac{n}{2}\log 2\pi e K
\nonumber \\
&\quad + I(T_1^n ;\sqrt{K}X^n+\sqrt{K}Z^n+
T_1^n|T_2^n\ldots T_K^n,W)\label{eq:Kuser_MutInf}\\
&\ge \frac{n(K-1)}{2}\log 2\pi e Q + \frac{n}{2}\log 2\pi e K
+\left[\frac{n}{2}\log\left(\frac{Q}{K(P+1)}\right)\right]^+
\label{eq:KuserRDBound}
\end{align}

The justification for the above steps is as follows.
In~\eqref{eq:Kuser_Ortho} we have use the fact that the
differential entropy is invariant to a rotation,
while~\eqref{eq:Kuser_claim} follows from
Claim~\ref{claim:KuserGaussClaim}.
In~\eqref{eq:Kuser_Independence} and~\eqref{eq:Kuser_IID} we have
used the fact that $T_j^n$ are mutually independent, i.i.d.\ and
independent of $W$. Eq.~\eqref{eq:Kuser_MutInf} follows by
additionally conditioning the entropy term
in~\eqref{eq:Kuser_MutInf_ID} with $X^n$  and using the fact that
$Z^n$ is independent of $(W,X^n,T_1^n,\ldots T_K^n)$.
Finally~\eqref{eq:KuserRDBound} follows from fact that since
$T_1^n$ is independent of $\{T_j^n\}_{j=2}^K$ and $W$ we can use
an argument analogous to that in Lemma~\ref{eq:LemmaRD} to have
$I(T_1^n ;\sqrt{K}X^n+\sqrt{K}Z^n+ T_1^n|T_2^n\ldots T_K^n,W) \ge
\left[\frac{n}{2}\log\left(\frac{Q}{K(P+1)}\right)\right]^+ $.
Finally, substituting~\eqref{eq:KuserRDBound}
in~\eqref{eq:commonRateTrick_Kuser}, we
obtain~\eqref{eq:gauss_K_user_UB}.
\section{Proof of Upper Bound \protect\eqref{eq:gub_s} in Theorem~\protect\ref{thm:gaussbounds}}
\label{app:d}


Our proof is structured as follows. We derive an upper bound for a
particular single-interference Gaussian channel, and reason that
the capacity of the two-interference channel of interest in
Theorem \ref{thm:gaussbounds} cannot be higher.

\begin{figure}
\begin{center}
\psfrag{&1}{$W$} \psfrag{&2}{$S^n$} \psfrag{&3}{$X^n$}
\psfrag{&4a}{$Z_2^n$}\psfrag{&4b}{$Z_1^n$} \psfrag{&5a}{$Y_2^n$}
\psfrag{&5b}{$Y_1^n$}\psfrag{&6b}{$\hat{W}_1$}\psfrag{&6a}{$\hat{W}_2$}
\includegraphics[width = 3.5in, angle =0]{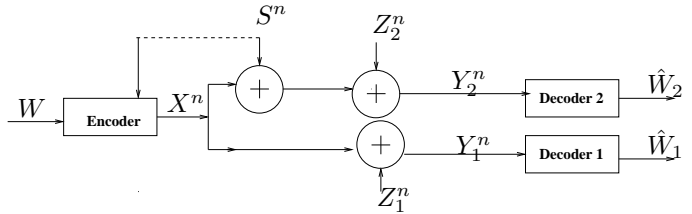}
\caption{Two-user Gaussian Channel with one-interference
sequences. We derive upper bound on the capacity of this channel
and show that this is also an upper bound for the two-interference
channel in Fig.~\ref{fig:gaussian}.  Here only receiver 2
experiences additive white Gaussian interference of variance $Q$.}
\label{fig:gaussian_one_interf}
\end{center}
\end{figure}

As shown in Figure~\ref{fig:gaussian_one_interf}, the  single-interference channel
is one in which $S_1^n=0$ and $S_2^n=S^n$. Only the second receiver experiences interference.

The subsequent two Lemmas establish that an upper bound on the
capacity of the single interference channel is also an upper bound
on the capacity of the two-interference channel in
Figure~\ref{fig:gaussian}.

\begin{lemma}
Suppose that for the single interference channel model in
Figure~\ref{fig:gaussian_one_interf}, the encoder and decoder 1
have access to a source of common randomness $\Theta$, which is
independent of the message $W$ and $(S,Z_1,Z_2)$. Then the
capacity of the single interference Gaussian channel is at-least
as large as the channel with two independent interferences in
Figure~\ref{fig:gaussian}. \label{lem:simulation}
\end{lemma}
\begin{proof}
The proof follows by observing that using the source of common
randomness $\Theta$, we can generate an i.i.d.\ Gaussian
$\cN(0,Q)$ sequence $S_C^n$, for any value of $n$. This sequence
is independent of all other channel parameters and is known to
both the encoder and decoder 1. It is used to simulate the two
independent interference channel as follows. Decoder 1, simply
adds this sequence to the received output, and ignores its
knowledge in decoding. The encoder has to deal with two sequences
$(S_C^n,S^n)$, both i.i.d.\ Gaussian $\cN(0,Q)$. With this
transformation, any coding scheme for the two interference channel
in Figure~\ref{fig:gaussian} can be used over this channel with
arbitrarily small probability of error.
\end{proof}

\begin{lemma}
A source of common randomness $\Theta$, which is independent of
the message $W$ and the channel parameters $(S,Z_1,Z_2)$ cannot
increase the capacity of the single interference channel in
Figure~\ref{fig:gaussian_one_interf}. \label{lem:CR}
\end{lemma}

\begin{proof}
Our proof is analogous to the proof that common randomness does
not increase the capacity in the single-user case in
\cite{ErezShamaiZamir01}. We argue that for any sequence of codes,
given a stochastic encoder and decoder that depends on the shared
random variable $\Theta$, there exists a deterministic encoder and
decoder with a smaller probability of error.

Given the message $m$ and state sequence $s^n$, and a realization
$\theta$ of the shared random variable, the encoding function
(c.f. Definition~\ref{def:channel}) be given by $x^n =
f(m,s^n,\theta)$. Similarly the decoding functions are given by
$\hat{m}_k = g_k(y_k^n,\theta)$ for $k =1,2,\ldots,K$. The average
probability of error for the rate $R$  randomized code is  then
defined by
\begin{equation*}
\begin{aligned} & P_e^{n,\mrm{randomized}} \\
&= \frac{1}{2^{nR}}\sum_{m=1}^{2^{nR}}
E_\Theta \left[\sum_{y^n: \exists k: g_k(y_k^n,\theta) \neq m} \sum_{s^n}
p(s^n) p(y^n|f(m,s^n,\theta)) \right]\\
&= E_\Theta \left[\frac{1}{2^{nR}}\sum_{m=1}^{2^{nR}}  \sum_{y^n:
\exists k: g_k(y_k^n,\theta) \neq m} \sum_{s^n} p(s^n)
p(y^n|f(m,s^n,\theta)) \right]\\
 &= E_\Theta \left[\Pr\left\{\left.\bigcup_{k=1}^K \{
g(Y_k^n,\theta)\neq W \}\right\}  \right|  \Theta=\theta\right],
\end{aligned}
\end{equation*}
where the second equality follows by interchanging the expectation
and summation over $m$, and the third equality follows by
observing that given a realization of the random variable $\Theta$, the
encoding and decoding are both deterministic and we can use the
definition of the average probability of error in
\eqref{eq:errprob}. Finally note that there must be some value of
$\theta$ for which the term inside the expectation is minimized. We can
design the encoding and decoding function for this deterministic
value of $\theta$ and our probability of error will be lower than the
average. Thus having access to common randomness cannot decrease
the probability of error for the channel of interest.
\end{proof}

Lemma~\ref{lem:simulation} and~\ref{lem:CR} imply that an upper bound on the capacity of
the single interference channel in Figure~\ref{fig:gaussian_one_interf} is also an upper bound
on the two independent-interference channel in Figure~\ref{fig:gaussian}. So we will derive an upper
bound for the former.

Invoking the result of Lemma \ref{lem:CorrelDoesNotMAtter}, we can
let $E[Z_1 Z_2] = \rho$, where $\rho \in [-1,1]$ will be optimized
later. As in the previous Appendix define $Z_- \defeq
(Z_1-Z_2)/\sqrt{2}$ and $Z_+ \defeq (Z_1 + Z_2)/\sqrt{2}$.

Suppose $R_1$ and $R_2$ denote the rates at which the two
receivers can reliably decode the common message. The rate of the
common message must satisfy $R \leq \min(R_1,R_2)$. Similar to our
derivation in Appendix~\ref{app:c}, we use Fano's inequality to
bound $R_1$ and $R_2$ as
\begin{align}
nR_1 &\le \frac{n}{2}\log 2\pi
e(P+1)-h(Y_1^n|W)+n\epsilon_n,\label{eq:g-conv-r1-2}\\
nR_2 &\leq \frac{n}{2} \log 2\pi e(P+1+Q +
2\sqrt{PQ})-h(Y_2^n|W)+n\epsilon_n.\label{eq:g-conv-r2-2}
\end{align}

Our bound for $R$ follows the derivation analogous to that
for~\eqref{eq:commonRateTrick} and is given by
\begin{equation}
\begin{aligned}
nR &\le \frac{n}{4} \log 2\pi e(P+1+Q + 2\sqrt{PQ}) \\
&\quad + \frac{n}{4}\log 2\pi e(P+1)-\frac{1}{2}h(Y_1^n,Y_2^n|W) +
2n\eps_n
\end{aligned}
\label{eq:commonRateTrick2}
\end{equation}

It remains to lower bound the joint-entropy term
in~\eqref{eq:commonRateTrick2}.

\begin{align}
&\quad h(Y_1^n,Y_2^n) \nonumber\\
&=
h\left(\frac{Y_1^n+Y_2^n}{\sqrt{2}},\frac{Y_1^n-Y_2^n}{\sqrt{2}}\Biggm| W\right) \label{eq:Orth_trans}\\
&= h\left(\sqrt{2}X^n + Z_+^n + \frac{1}{\sqrt{2}}S^n,
 -\frac{1}{\sqrt{2}}S^n+Z_-^n \Biggm| W \right)\nonumber\\
&= h\left(-\frac{1}{\sqrt{2}}S^n+Z_-^n \Biggm| W\right)\nonumber\\
&\quad+ h\left(\sqrt{2}X^n + Z_+^n + \frac{1}{\sqrt{2}}S^n\Biggm|
W,-\frac{1}{\sqrt{2}}S^n+Z_-^n \right) \nonumber\\
&\ge h\left(-\frac{1}{\sqrt{2}}S^n+Z_-^n\right)
\nonumber\\
&\quad+h\left(\sqrt{2}X^n + Z_+^n + \frac{1}{\sqrt{2}}S^n\Biggm|
W,-\frac{1}{\sqrt{2}}S^n+Z_-^n,S^n,X^n \right)
\label{eq:indAndCond}\\
&= h\left(-\frac{1}{\sqrt{2}}S^n+Z_-^n\right)+h(Z_+^n) \label{eq:indAndCond2}\\
&= \frac{n}{2}\log 2\pi e\left(\frac{Q}{2}+1-\rho\right)+
\frac{n}{2}\log 2\pi e\left(1+\rho\right) \label{eq:hy1y2bound2}
\end{align}
In the above steps,~\eqref{eq:Orth_trans} follows from the fact
that differential transformation is invariant under a pure
rotation,~\eqref{eq:indAndCond} follows from the fact that the
pair $(S^n,Z_-^n)$ is independent of $W$ and conditioning on
additional terms only reduces the second term,
while~\eqref{eq:indAndCond2} is follows from the fact that $Z_+^n$
is independent of all other variables in the second term.

Substituting~\eqref{eq:hy1y2bound2}
into~\eqref{eq:commonRateTrick2} and rearranging, we get
\begin{equation}
\begin{aligned}
R &\leq \frac{1}{4}\log\left(\frac{1+P}{1+\rho}\right) \\
&\quad + \frac{1}{4}\log\left(\frac{P+Q+1+
2\sqrt{PQ}}{Q/2+1-\rho}\right) + \eps_n, \\
\end{aligned}
\label{eq:upperbound-rho}
\end{equation}

Thus we have shown the expression for~\eqref{eq:gub_s_opt}. To
obtain the tightest bound we minimize the right hand side of the
above over $\rho$. The tightest bounds is obtained with the choice
\begin{equation}
\rho^*(Q) =
\begin{cases}
Q/4 & \text{ if } Q \leq 4 \\
1   & \text{ if } Q > 4.
\end{cases}
\label{eq:opt-rho}
\end{equation}
Substituting this value of $\rho$, in~\eqref{eq:upperbound-rho}
yields~\eqref{eq:gub_s}.

\subsection{Gains from Feedback}
\label{app:d-Feedback Argument} As noted in
Appendix~\ref{app:c-Feedback Argument},in the presence of causal
feedback it still holds that $Z_{+,i}$ is independent of
$(W,Z_{-}^n,S^n,X_1^i)$. It can be verified that with this
condition, the derivation that leads to~\eqref{eq:hy1y2bound2}
continues to hold and the upper bound in~\eqref{eq:upperbound-rho}
remains valid. One cannot however optimize over $\rho$ in the
presence of feedback as Lemma~\ref{lem:CorrelDoesNotMAtter} fails
to hold in the presence of feedback.

\section{Case of Correlated Interferences}
\label{appe-sec:corrInterf} In this section, we present the
derivation of the upper bound in
Theorem~\ref{thm:CorrelatedInterf}. The derivation is a minor
modification of the derivation for the case of independent
interferences. So only the steps that need to be modified will be
presented. As in the statement of the Theorem, we assume that
$S_1\sim \cN(0,Q_1)$, $S_2\sim\cN(0,Q_2)$ and $S_1-S_2 \sim
\cN(0,Q_d)$.

We first note that using Fano's inequality and the steps that lead
to~\eqref{eq:commonRateTrick} in Appendix~\ref{app:c}, an upper
bound on the common rate can be shown to be
\begin{equation}
nR \le \frac{1}{2}h(Y_1^n)+\frac{1}{2}h(Y_2^n)-
\frac{1}{2}h(Y_1^n,Y_2^n|W) + n\eps_n
\label{eq:commonRateTrick_correlated}
\end{equation}
Using the power constraint, we upper bound $h(Y_i^n) \le
\frac{n}{2}\log 2\pi e(P+Q_i+1+2\sqrt{PQ_i})$ for $i=1,2$. It
remains to lower bound the joint entropy term. In what follows, we
denote $Z_+^n = \frac{Z_1^n+Z_2^n}{2}$ and $Z_-^n = Z_1^n-Z_-^n$.
Note that $Z_+^n$ and $Z_-^n$ are mutually independent and i.i.d.\
samples from $\cN(0,(1+\rho)/2)$ and $\cN(0,2(1-\rho))$
respectively.
\begin{align}
&h(Y_1^n,Y_2^n|W) = h\left(Y_1^n-Y_2^n,
\frac{Y_1^n+Y_2^n}{2}|W\right)
\label{eq:det_trans}\\
&= h\left(S_1^n-S_2^n+Z_-^n, X^n+ \frac{S_1^n+S_2^n}{2} + Z_+^n|W\right)\nonumber\\
&= h(S_1^n-S_2^n+Z_-^n) + h\left(X^n+ \frac{S_1^n+S_2^n}{2} + Z_+^n|W\right) \label{eq:independence}\\
&\ge h(S_1^n-S_2^n+Z_-^n)+ h(Z_+^n)\label{eq:independence_of_Z^n}\\
&= \frac{n}{2}\log 2\pi e (Q_d + 2(1-\rho)) + \frac{n}{2}\log 2\pi
e\left(\frac{1+\rho}{2}\right)\nonumber
\end{align}
Here $\eqref{eq:det_trans}$ follows from the fact that the
transformation $\left[%
\begin{array}{cc}
  1 & -1 \\
  1/2 & 1/2 \\
\end{array}%
\right]$ has unit determinant and the differential entropy is
invariant to this transformation,~\eqref{eq:independence} from the
fact that $S_1^n-S_2^n+Z_-^n$ is independent of $W$ and
~\eqref{eq:independence_of_Z^n} from the fact that $Z_+^n$ is
independent of all other variables. The optimal value of $\rho$,
which yields the largest value for the lower bound is given by
$\rho^* = \min(1,Q_d/4)$ and the corresponding lower bound is
given by:
\begin{equation}
h(Y_1^n,Y_2^n) \ge
\begin{cases}
n\log (2\pi e)^2\left(1+\frac{Q_d}{4}\right) & \text{ if } Q_d \leq 4 \\
\frac{n}{2}\log (2\pi e)^2 Q_d     & \text{ if } Q_d > 4.
\end{cases}
\label{eq:hy1y2bound_corr_lb}
\end{equation}
Finally substituting~\eqref{eq:hy1y2bound_corr_lb}
in~\eqref{eq:commonRateTrick_correlated} gives us the expression
in~\eqref{eq:CorrelatedInter_UB}.

\section*{Acknowledgement}
The authors thank two anonymous reviewers for their insightful
comments which helped to improve the quality of the paper.

\bibliography{main}
\bibliographystyle{IEEEtranS}



%






\end{document}